\documentclass[10pt,journal,compsoc]{IEEEtran}
\ifCLASSOPTIONcompsoc
  \usepackage[nocompress]{cite}
\else
  \usepackage{cite}
\fi

\ifCLASSINFOpdf
   \usepackage[pdftex]{graphicx}
   \graphicspath{{../pdf/}{../jpeg/}}
   \DeclareGraphicsExtensions{.pdf,.jpeg,.png}
\else
   \usepackage[dvips]{graphicx}
   \graphicspath{{../eps/}}
   \DeclareGraphicsExtensions{.eps}
\fi

\usepackage{graphicx}
\usepackage{comment}
\usepackage{multirow} 
\usepackage[table,xcdraw]{xcolor}
\usepackage{subfigure}
\usepackage{epstopdf}
\usepackage{amsmath}
\usepackage{array}
\usepackage{tikz}
\usepackage{todonotes}

\usepackage{array}

\begin{document}
%
\title{Measuring the Effects of Scalar and Spherical Colormaps on Ensembles of DMRI Tubes}

\author{
Jian~Chen,~\IEEEmembership{Member,~IEEE},
Guohao~Zhang,~\IEEEmembership{Student Member,~IEEE}, \\
Wesley~Chiou,
David~H.~Laidlaw,~\IEEEmembership{Fellow,~IEEE}, 
and
Alexander~P.~Auchus 
\IEEEcompsocitemizethanks{
\IEEEcompsocthanksitem J. Chen is with the Computer Science and Engineering Department, 
The Ohio State University, OH 43210. E-mail: chen.8028@osu.edu.
\IEEEcompsocthanksitem G. Zhang and W. Chiou are with the Department
of Computer Science and Electrical Engineering, University of Maryland, Baltimore County, MD
 21025.
E-mail: \{guohaozhang, wchiou1\}@umbc.edu. 
\IEEEcompsocthanksitem D.H. Laidlaw is with Computer Science Department, Brown University, RI 02912. E-mail: dhl@cs.brown.edu.
\IEEEcompsocthanksitem A.P. Auchus is with the Neurology Department at the University
of Mississippi Medical Center, Jackson, MS 39216. E-mail: aauchus@umc.edu.
}
}

\markboth{Submitted to Transactions on Visualization and Computer Graphics,~Vol.~X, No.~X, December~2017}%
{Chen \MakeLowercase{\textit{et al.}}: }

\IEEEtitleabstractindextext{%
\begin{abstract}

We report empirical study results on the color encoding of ensemble scalar and orientation to visualize  diffusion magnetic resonance imaging (DMRI) tubes. 
The experiment tested six scalar colormaps for average fractional anisotropy (FA) tasks (grayscale, blackbody, diverging,  isoluminant-rainbow, extended-blackbody, and coolwarm)  and four three-dimensional (3D) directional  encodings for tract tracing tasks (uniform gray, absolute, eigenmap, and Boy's surface embedding). We found that extended-blackbody,
coolwarm, and blackbody remain the best  three approaches for identifying ensemble average in 3D.
Isoluminant-rainbow coloring led to the same ensemble mean accuracy as other  colormaps. However,  more than $50\%$ of the answers consistently had higher estimates of the ensemble average, independent of the mean values. 
Hue, not luminance, influences ensemble estimates of mean values. 
For ensemble orientation-tracing tasks, we found that the Boy's surface embedding (greatest spatial resolution and contrast) and absolute color (lowest spatial resolution and contrast) schemes led to more accurate answers than the eigenmaps scheme (medium resolution and contrast), acting as the uncanny-valley phenomenon of visualization design in terms of accuracy.

\end{abstract}

}

\maketitle

\IEEEdisplaynontitleabstractindextext
\IEEEpeerreviewmaketitle

\IEEEraisesectionheading{\section{Introduction}\label{sec:introduction}}


\IEEEPARstart{E}{xploratory} 
vector and tensor field visualizations studying regions of interest or a group of objects at a time~\cite{phadke2012exploring} count on the human visual system to extract statistical information from features. Perceiving average or other statistical features from a group of similar items,
called \textit{ensemble perception}~\cite{leib2016fast}~\cite{chetverikov2017representing}, is a robust visual phenomenon studied largely in vision science that operates across a host of visual dimensions: size~\cite{ariely2001seeing},
orientation~\cite{robitaille2011more},
position~\cite{alvarez2008representation}, 
motion~\cite{williams1984coherent}, speed~\cite{watamaniuk1992human}, number~\cite{burr2008visual}, identities~\cite{neumann2013viewers}, structures~\cite{oliva2006building}, and luminance~\cite{bauer2009does}.


The applicability of these vision science results to visualizations is anecdotal because of at least two methodological 
barriers between these two domains.
Vision science studies are intended to capture static views, separate perception and cognition from interaction, and also separate domain-specific uses from visual stimuli.
In contrast, in visual exploration these factors must be integrated. Additionally, 
spatial visualization features, such as continuity,  symmetry, and clusters, may not be present in images.

\begin{figure*}[t!]
\centering
  \includegraphics[width=\linewidth]{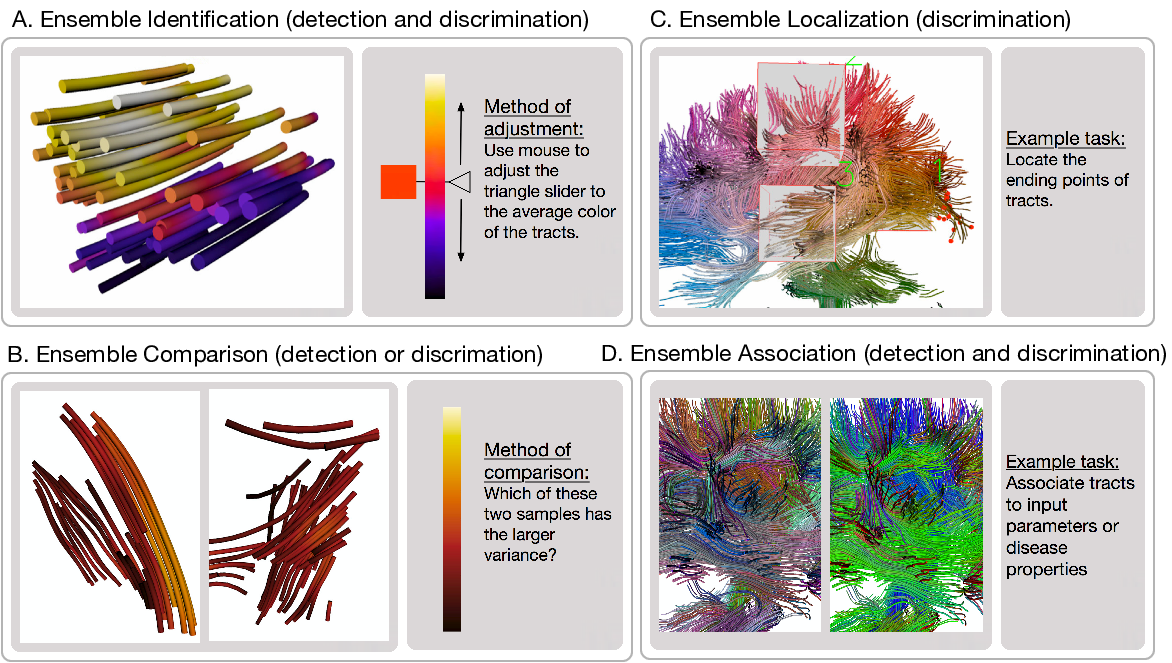}%
\caption{
Four Types of Ensemble Tasks and Example Methods for Testing Ensemble Visualization
}
\label{fig:method}
\end{figure*}

Working in collaboration with brain scientists, we have recognized two main challenges for showing 3D diffusion magnetic resonance imaging (DMRI) tractography. The first is to support ensemble univariate representations. Scalars are commonly encoded in one-dimensional (1D) colormaps, e.g., showing fractional anisotropy (FA) measured at every voxel to quantify disease states~\cite{chua2008diffusion}. Though univariate coloring has been extensively studied in two-dimensional (2D) data visualizations (see the excellent reviews by Zhou and Hansen~\cite{zhou2016survey} and Silva et al.~\cite{silva2011using}), 3D color ensembles may introduce constraints in three respects.
First, the univariate schemes of luminance and hue combination that work well in 2D may not apply in 3D, since luminance contrast can belie color constancy and distort 3D shape perception due to lighting~\cite{finlayson2001color}. Second, shading prevents the use of dark colors~\cite{kim2009modeling}~\cite{moreland2016we}, thus reducing the number of differentiable color s.png. Third, interpreting ensembles may not require visually deriving individual values~\cite{chetverikov2017representing}. Since scientific data are often continuous, the human visual system may well optimize strategies for efficient visual detection~\cite{christen2013colorful}.


The second challenge is showing spatial brain structural connectivities from tracts (often rendered as tubes). This task requires the viewer to visually segment collections of tracts of various directions. Szafir et al. call this type of task  \textit{ensemble subset extraction}~\cite{szafir2016four}; Phadke et al. call it \textit{attribute value exploration}~\cite{phadke2012exploring}. DMRI tracts, unlike data in these studies are continuous in space and carry domain-specific attributes such as \textit{symmetry} and 
\textit{proximate regions}~\cite{pajevic1999color}~\cite{demiralp2009coloring}. 
As a result, novel tract colorings concerning \textit{tract locality}, \textit{angular uniformity}, and \textit{spatial resolution} have been explored~\cite{demiralp2009coloring}. No design knowledge exists, however, to quantify the practicality of these spherical colormaps in visualization.


The present work addresses these two important challenges by first summarizing a set of ensemble tasks of 
\textit{identification}, \textit{localization},  \textit{comparison}, and \textit{association} (Fig.~\ref{fig:method}). We then examine two identification
tasks by evaluating state-of-the-art coloring methods. 
Specifically, we answer the following questions: 
\textit{How reliable are colormaps for deriving ensemble averages from 3D spatially distributed tracts?} 
\textit{Which colormaps are applicable to ensemble average?} 
\textit{Which is the most effective ensemble orientation extraction technique?}

Our  work  makes  the following contributions. 

\begin{itemize}

\item 
Formally proposes and expands  ensemble visualization concepts inspired by vision science.

\item
Establishes new measurement metrics  that  permit us to match  the  data  characteristics to color  characteristics  by analyzing data  distributions.

\item
Suggests practical ensemble quantification methods to characterize visual dimensions. 

\item 
Derives some design recommendations for spatially continuous datasets for ensemble average and orientation discrimination.

\end{itemize}

\section{Terms and Related Work}

Our work draws upon work related to (1) ensemble color representation results in vision science and (2) univariate and orientation representations in visualization.
In this section, we  broaden the definition of ensemble encoding in visualization 
and connect color theory to ensembles and relevant study results. 


\subsection{Ensemble Encodings: Definition}

The \textit{ensemble} concept in visualization often refers to a collection of datasets and is perhaps best known as ensemble simulation and uncertainty quantifications~\cite{potter2009ensemble}. Ensemble has been broadly studied in vision science (e.g.,{~\cite{whitney2017ensemble}}), where \textit{ensemble representation} is used to explore how humans use statistical regularities in a group of similar objects to process information{~\cite{alvarez2011representing}}.



Our current work  supports this recent  broad perspective on the role of visual statistical processing and  embraces the  idea  that  these  visualization tasks,  whether from  ensemble simulations (statistical  properties such as uncertainty{~\cite{potter2012quantification}}{~\cite{sanyal2010noodles}}) or not (e.g., overviews and  detecting global  features in  flow  fields{~\cite{laramee2004state}} and   areas   
or sets{~\cite{max1990area}}), share the property that multiple measurements are combined to give rise to a higher-level statistical description.

Following this new human information processing perspective, we formalize \textbf{ensemble representation}
as an umbrella term encompassing existing 3D visualization methods that demand the human visual system to derive statistical attributes from data. 
For example, correlated textures along vector fields help humans derive ensemble patterns to see flow movement; glyphs enable efficient visual assessment  of  ``a chunk of flow''~\cite{forsberg2009comparing}. 

\subsection{Color Ensembles}

The study of \textbf{color ensemble representations} concerns how our visual system derives statistical information through visual processing of color features. Human vision can effectively process color ensembles, for example, to discount spectral variations and assign stable colors to objects to achieve consistent scene representation~\cite{haberman2012ensemble} and color constancy~\cite{finlayson2001color}. 
Color ensembles facilitate scalable visual inspection. 
Mauly  and  Franklin{~\cite{maule2015effects}}
study a series of uniformly colored  circular elements ranging from
4 to  48 items  subtended at  12, 20, and  28 just-noticeable differences (JNDs) and  report that the accuracy was insensitive to changes in the number of elements  in an ensemble. Only reaction time was longer for ensembles with more hues.
Ensemble processing does not require focused attention to subsample of elements~\cite{bauer2009does}. 

Dedicated human visual processing of ensemble colors may also exist~\cite{haberman2015individual}.
In a 2D time-varying chart visualization, Correll et al. find that color is more effective than position for showing averages and distributions~\cite{correll2012comparing}; this is contradictory to classical design recommendations in which position is more accurate than color for quantitative comparisons, when ensemble is not required~\cite{cleveland1984graphical}.
Also reliable average estimates can be made from two hues  of red-blue, blue-green, and yellow green{~\cite{webster2014perceiving}}  and categorical boundaries can be accurately labelled 
for greenish-blue and  bluish-green{~\cite{wright2011effects}} and gray-scale alike textures{~\cite{zhao2017bivariate}}.

These intriguing results on ensemble hues, mostly presented in vision science, seem to refute the idea that ensemble hues cannot be described in terms of magnitude but as qualitative experiences. They may be effective for ensemble averages and boundary detection when the data or hue variance is localized and small.
In this work, we chose several multihue colormaps, such as extended-blackbody and coolwarm. We also use a reasonably good rainbow colormap, \textit{aka} Kindlmann et al.'s isoluminant rainbow~\cite{kindlmann2002face}.  We compare these approaches against other univariate methods.

\subsection{Univariate Coloring}

The most influential color studies lie in univariate colormap design and characterizations (e.g., color harmony and categories~\cite{hu2014interactive}~\cite{gramazio2017colorgorical},
metrics~\cite{bujack2017good},
and modeling~\cite{szafir2017modeling}).
Silva et al.~\cite{silva2011using}
and Zhou and Hansen~\cite{zhou2016survey} 
summarize color characteristics important in univariate colormap design,  such as 
\textit{ordering} (colormaps  must  preserve  the order in data), 
\textit{separation} (different data must be perceived differently)~\cite{trumbo1981theory}, and 
\textit{uniformity} (perceived differences in color must accurately reflect numerical data differences). 
Among these characteristics, uniformity is believed most important{~\cite{ware2017evaluating}}. 
Rainbow colormap is believed to be poor at showing quantitative data because it lacks nearly all these attributes.

This design knowledge led us to adopt several univariate maps suggested by Moreland~\cite{moreland2016we}, including extended-blackbody (monotonic luminance and multihue), blackbody (perceptually uniform, monotonic luminance and multihue), coolwarm (perceptually uniform, two-hues and monotonic on each side), and diverging (two-hues and perceptually uniform and monotonic on each side). Some of them have also been incorporated in the popular 3D visualization
tools VTK and Paraview. 
Since color ensembles by hues might be effective,
we have also used  Kindlmann et al's isoluminant rainbow{~\cite{kindlmann2002face}} and used gray-scale as the baseline methods. 

\subsection{Vector and Tensor Field Evaluation}

Pioneering 3D vector  and  tensor  field  studies have  largely focused on univariate comparisons, such as vector speed between two locations~\cite{forsberg2009comparing}, tracing a single  tract~\cite{penney2012effects}, reading quantities at  each  sampling site~\cite{zhao2017validation}, and  showing depth and  distances between adjacent occluded tracts~\cite{ritter2006real}~\cite{svetachov2010dti}. An exception is the study by Acevedo and  Laidlaw~\cite{acevedo2006subjective} in which  participants were to discriminate boundaries through a set of size-varying circles and must visually derive groups from visualization.


Borkin et al.'s work~\cite{borkin2011evaluation}  closely resembles ours in terms of colormap comparisons to support seeing in 3D. That study compares rainbow and diverging colormaps for detecting regions of heart diseases after projecting 3D artery flow patterns to 2D and finds that a rainbow colormap decreases detection rates~\cite{borkin2011evaluation}. The present work builds on these studies but expands the scope in two important ways: we measure more tasks to understand ensemble averages and
direction discrimination, and our tasks are in 3D. 
We further formalize  the task space in Chen et al.~\cite{chen2012effects} for ensemble 
univariate and spherical direction discrimination.

\subsection{Continuous Ensemble Spherical Colormaps}


Knowledge about effective spherical colormap design is limited, despite their importance for showing tensor and vector fields.
To show brain connectivity through tracts, 
Pajevic  and Pierpaoli~\cite{pajevic1999color} use elegant solutions through extensive studies on \textit{rotation} and \textit{mirror symmetry}. The absolute values of the $xyz$-coordinates of the principal diffusion tensor eigenvectors are mapped directly to RGB color-triples. The advantages of this \textit{absolute} approach include: (1) perceptual uniformity, (2) user familiarity with RGB colors associated with a vertebrate direction, (3) high contrast between vertebrate directions, and (4) four-way symmetry (left-right, dorsal-ventral, anterior-posterior, and antipodal.)
Even though this absolute encoding approach provides a seemingly low-resolution view of tract orientation, our brain scientist collaborators suggest that this colormap dominates brain science because it conveys most important transverse, sagittal, and coronal directions.


Other solutions reveal patterns and increase spatial resolutions. Kindlmann et al. introduced a \textit{hue-ball} approach and a barycentric map for direct volume rendering of tensor fields by assigning color and opacity based on the direction of the principal eigenvector and anisotropy type of the diffusion tensor~\cite{kindlmann1999hue}. An attractive characteristic of this approach is its high contrast between adjacent tracts: they are colored with bright, saturated colors spanning from red, yellow, green, cyan, blue to purple. Demiralp et al.~\cite{demiralp2009coloring}  use \textit{Boy's real projective plane immersion} to visualize the direction of brain tracts. This \textit{Boy's surface} coloring possesses good \textit{locality} and \textit{contrast} by showing the finest details, and has the greatest \textit{spatial resolution} of all spherical colormaps.

Vision science has studied multihue mainly as a pattern-segmentation mechanism for identifying structural variations. 
Maule  et al.{~\cite{maule2014getting}} suggest that there  
may be a functional limit to the amount of variance that can be rapidly encoded by summary statistics of set discriminations.
Such set discriminations, though close to our orientation discrimination, can prescribe methods only for discrete clusters. 
No study exists to our knowledge to explore to what extent continuous spherical coloring of ensemble line field would be most beneficial. 
Our study compares four techniques to understand the effectiveness of ensemble orientation discriminations. 
Our hypothesis is largely driven by the
vision science literature positioning that colormaps with higher resolution could improve the spherical color direction detection.

\section{Brain DMRI Data Characterization and Ensemble Tasks}

This section first describes the data and task characterization 
by following Munzner's{~\cite{munzner2009nested}} data and task abstraction method,
and then presents our measurement method.

\subsection{Brain DMRI Data Characterization} 

DMRI measures water diffusion as a second-order positive-definite tensor~\cite{mori2006principles}. 
Water diffusion patterns have been analyzed comprehensively by brain scientists to study anatomical fibrous structures. 
Modern advances have  extended to meta-analysis  of  brain cohorts~\cite{zhang2017overview}. 
Visualization design guidelines for understanding complex spatial structures have also been a recent focus~\cite{isenberg2013systematic}, albeit disproportionately small in the amount of empirical work directly focused on evaluation. 
Preim et al.{~\cite{preim2016survey}} have surveyed perceptually-motivated 3D visualization for medical imaging visualization, but focused 
on depth and shading.
This challenge in coloring MRI datasets is often cited as a top visualization challenge~\cite{christen2013colorful}.

The first and most reliable benchmark measurement is 
fractional anisotropy (FA)~\cite{kochunov2016heterochronicity}. FA, a normalized scalar, measures the water diffusion patterns: a value of zero means that diffusion is isotropic, i.e., it is unrestricted in all directions (usually in gray matter); a value of one means that diffusion occurs only along one axis and is fully restricted (usually in white matter).  Brain scientists are concerned with average FAs in regions containing a set of voxels or tracts. 
In this study, FAs are in the range of [0.2, 1] and average FAs are [0.25, 0.85].
 

Another important measurement is brain structural connectivities~\cite{chen2012effects}.
A continuous diffusion tensor field is first constructed from the measured DMRI data. Tracts are then computed at voxel sampling  locations 
via tractography, a 3D technique for representing brain structural  connectivity{~\cite{zhang2003visualizing}}. We terminate 
tract tracing when the FA value is less than 0.2.
The tracts   are  depicted to show connectivity information.
A group of tracts sharing similar orientations is called a bundle. 
Some studies use template-based approaches to derive and color tracts to show anatomical connectivity; others attempt to visualize the structures independent of templates. Our current work studies
five major bundles labelled by our collaborators. 

Several brain analysis tools and methods have supported colormaps. For example, DTI Studio lets users manually assign selected tracts a color as well as use the default randomly assigned colors for individual  tracts~\cite{jiang2006dtistudio}. 
3D Slicer lets users select among a large variety of colormaps or customize their own for visualizing variables~\cite{pieper20043d}. While these tools offer great flexibility, our results can give users more informed design choices among techniques and tools.

\subsection{Ensemble Task Characterization}

\subsubsection{Four Task Categories}
\label{sec:tasks}

We obtain the following measurable low-level tasks (Fig.{~\ref{fig:method}}).  In each category, we separate detection (e.g., which is higher?) and discrimination (e.g., how much higher?) tasks
inspired by Borgo et al.{~\cite{borgo2014order}} and Zhao et al.{~\cite{zhao2017validation}}, so as to address the goal of design for perceptually accurate visualizations.

\begin{enumerate}
\item

\textit{Ensemble identification} is performed when the goal is to read mean values 
or estimate the probability distributions of values from \textit{similar} objects.

Some typical identification tasks are: what are the average FA values (Fig.{~\ref{fig:method}}(A))? Where is the boundary between regions of 
different anatomical structures? Do the two bundles belong to different groups?  What is the average brain?

\item
\textit{Ensemble comparison} is useful to compare multiple ensembles or items or identify the most common outputs.
Some example tasks are: are the left and the right hemispheres of CC different?  If so, by how much? 
The task in Fig.{~\ref{fig:method}}(B) compares between diseases
outcomes in cohorts.

\item
\textit{Ensemble localization} asks the viewer to find where a certain ensemble value or attribute is located within the data.
Fig.{~\ref{fig:method}}(C) stresses visual lookup and asks where the lesion  is in the brain.  Where  are regions of maximum and  minimum mean  FA values?

\item
\textit{Ensemble association} involves determining the  associative relationships  between  or among  \textit{related}  objects. 
Fig.{~\ref{fig:method}}(D) shows  the  average tracts  computed from ensembles.
Some example tasks  are: which  of these two  average brains  is  associated with dementia? And  at what  state  of the  dementia? 
Using  a simulator  and  after  varying parameter A, what  are  the associated brain regions sensitive to  these  inputs, and  what  is the distribution of the changes among these output ensembles?
\end{enumerate}

\subsection{Metric}

There  are  several  considerations in  measuring the  ensemble representations. We  divide the  data  or  the  colormap into bins  to  represent sub-regions. This  is because a region  of interest (ROI) in a spatial volume is likely  to be localized to a group of data  points. Also,  we  can  associate the  data distributions in each bin to color distributions in a colormap to understand colormap usefulness. For example, the spread or  variance of  the  resulting distribution in  each  bin  in  a colormap reflects  the  ensemble average performance. The shape  of the results also reflects the sensitivity of features or dimensions to the ensembles. Robust  sensitivity to summary statistics will  yield  a  narrow distribution. A function can also  be  fitted  to  the  data   to  reveal  sensitivity to  the discriminative threshold to measure accuracy. In this  work, for ensemble average we divide the input data into  12 bins  and  randomly sample the  data  such  that  each bin  has  a  high- fidelity   representation of  the  DMRI  tract attributes. For orientation detection tasks, we follow past practice and  measure the responses to spherical colormaps by randomly sampling the input.



\section{Ensemble Experiment for Brain DMRI Visualizations}

The objective here is to determine which  ensemble colormaps are  more  accurate for  showing DMRI  datasets. We  are  particularly interested in  the  first  task  type,  ensemble  identification (of mean  and  orientation) (Section{~\ref{sec:tasks}}).

\subsection{Hypotheses}

Given our own experiences, our collaborators' subjective choices, and the literature, 
we have had five hypotheses when entering the experiment:

\begin{itemize}

\item H1 (rainbow hypothesis). For  ensemble average, the isoluminant rainbow colormaps may be  as  accurate as uniform and  monotonic 
luminance colormaps.

\item H2 (multihue hypothesis). For  ensemble average, multiple-hue  colormaps can be effective  for reading ensemble mean  values. 

\item
H3 (gray hypothesis). For ensemble average, baseline gray   may   have   the worst accuracy  for  ensemble mean  of scalar values.

\item H4 (direction detection hypothesis).
For  ensemble orientation  detection, higher spatial resolution can improve orientation accuracy.

\item H5 (colorfulness hypothesis). Having color  is  better than  baseline  no-color  uniform representation for identifying orientations.

\end{itemize}

\subsection{Three-Dimensional Ensemble Tasks}
\label{sec:task}

\begin{figure}[t!]
\centering
\subfigure[Task 1: Ensemble Average] {\includegraphics[width=0.95\linewidth]{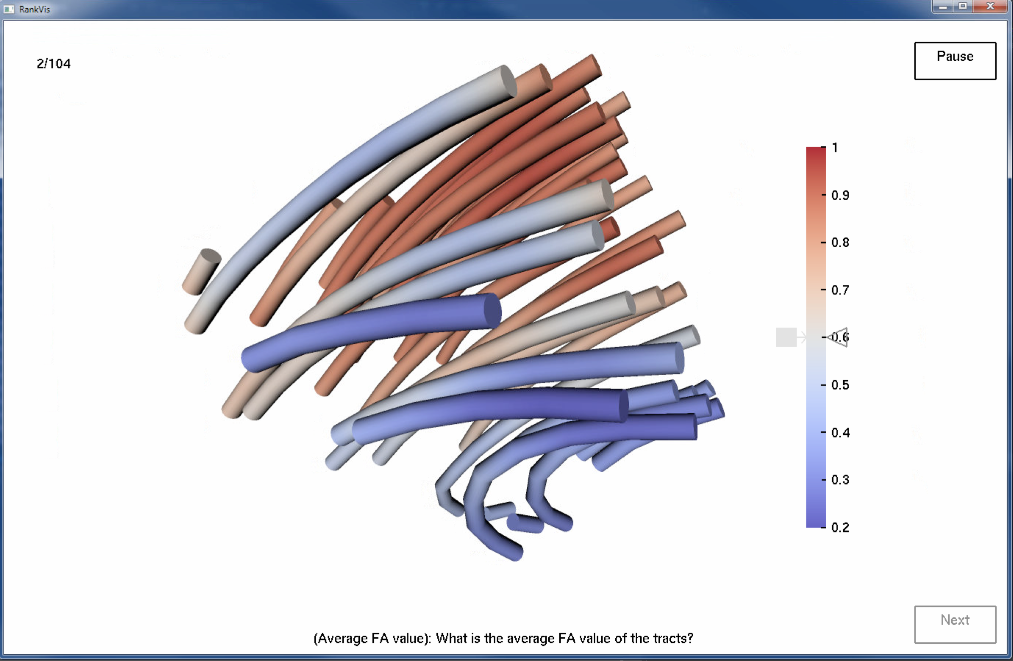}
\label{fig:task1:taskFA}}

\subfigure[Task 2: Ensemble Orientation]
{\includegraphics[width=0.95\linewidth]{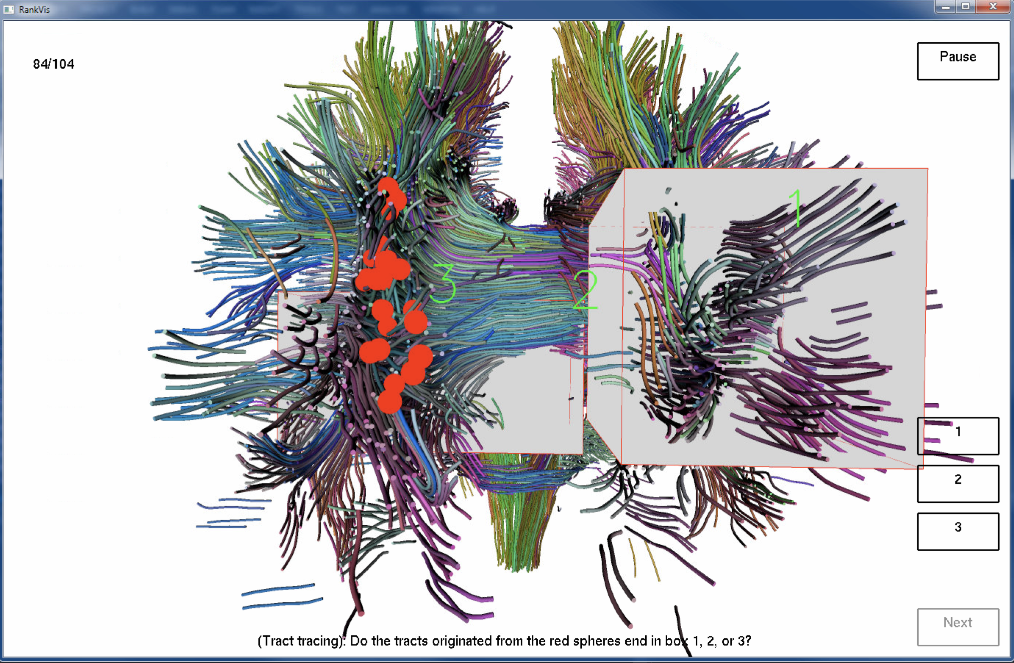}
\label{fig:task2:tracing}}

\caption{
Two Ensemble Identification Tasks in the Empirical Study. (a) What is the average value of the tracts? This example uses the diverging colormap. (b) Do the tubes originating from the red spheres end in box 1, 2 or 3? This example uses a Boy's surface colormap.
}
\label{fig:tasks}
\end{figure}

\subsubsection{Task 1: Ensemble Average (Discrimination task)}

Figure~\ref{fig:task1:taskFA} shows an example task in which participants were asked to label the average FA values of the brain areas sampled in a ROI. 
The participants indicate their answer for each task by dragging the slider on the screen to show the average color. 
The answers are evenly distributed along the 12 bins (see Section 5.3) so that participants are not biased.

\subsubsection{Task 2: Ensemble Orientation (Detection task)}

Figure~\ref{fig:task2:tracing} shows an example task in which participants were asked to find the one box among three in which the endpoints of the tracts marked by red spheres lay at one end of the tracts. Participants were told that the marked tracts in the same bundle followed the same orientation (anterior-posterior, dorsal-ventral, or left-right).


\subsection{Choosing Ensemble Colormaps}

\subsubsection{Six Univariate Colormaps for Ensemble Average}

\begin{figure}[!tp]
\centering
\includegraphics[width=\linewidth]{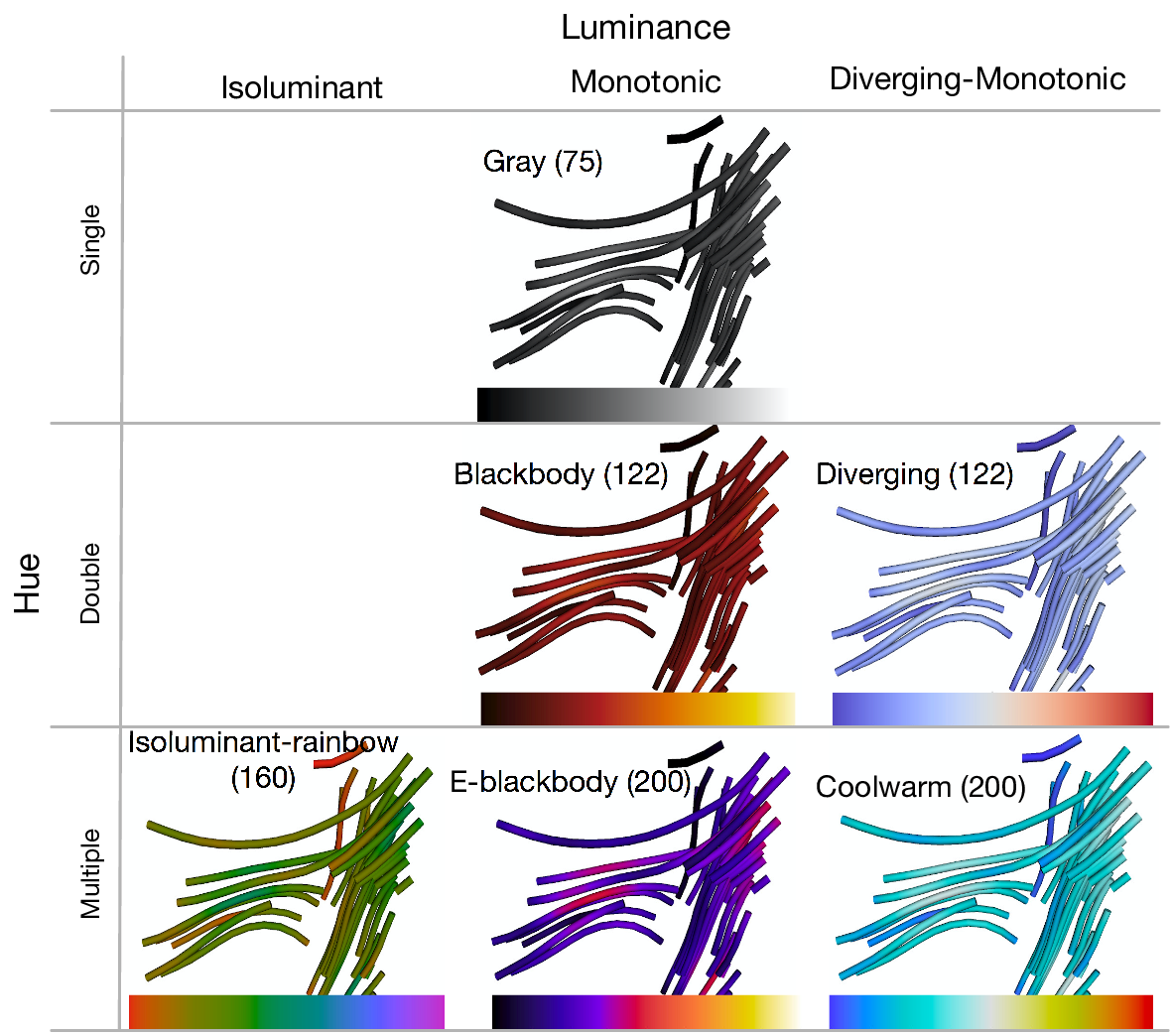}%
\caption{Task 1's Six Univariate Colormaps. The numbers after colormap names are arc-lengths in the L*A*B* color space.}
\label{fig:task1maps}
\end{figure}

Six univariate colormaps 
shown in Figure{~\ref{fig:task1maps}} are measured in task 1 (Ensemble Average). 

These colormaps are chosen due to their popularity, relevance to our hypotheses, and our collaborators' recommendations. 
Arc-length is computed with CIEDE 2000 by summing the DeltaE values along the curve in the L*A*B* color space{~\cite{sharma2005ciede2000}}.
All color interpolation is performed using linear interpolation in this L*A*B* color space. 
The dark part is cut out to keep these values as close as possible for each hue condition.
Appendix A shows the colormap profile in the {L*A*B*} color space.

The \textit{grayscale colormap} uses a single-hue and monotonic luminance with arc-length 75.

The \textit{blackbody colormap} is a double-hue and monotonic luminance map inspired by the wavelengths of light from blackbody radiation. 
We use arc-length 122 instead of 145 to match that of the diverging map.  We removed the dark end due to the low sensitivity to low luminance values.

The \textit{diverging} colormap contains two hues and increases/decreases luminance monotonically with arc-length 122. The closer the color is to the center of the color map, the higher the luminance.

The \textit{isoluminant-rainbow colormap} displays multihue rainbow with arc-length 160. It is isoluminant for the standard viewer,  
with the luminance level of 50. 

The \textit{extended blackbody colormap} is a monotonic luminance colormap and adds blue and purple hues to the blackbody map described above with arc-length 200.

The \textit{coolwarm colormap} has monotonically increased and decreased luminance. This colormap has the same luminance range and variations along the luminance direction as the diverging map; it adds yellow and cyan hues to the diverging map because these two hues are common transitions in coolwarm colormaps that use red and blue.

\subsubsection{Four Ensemble Direction Colormaps}

\begin{figure}[!tp]
\centering
\includegraphics[width=\linewidth]{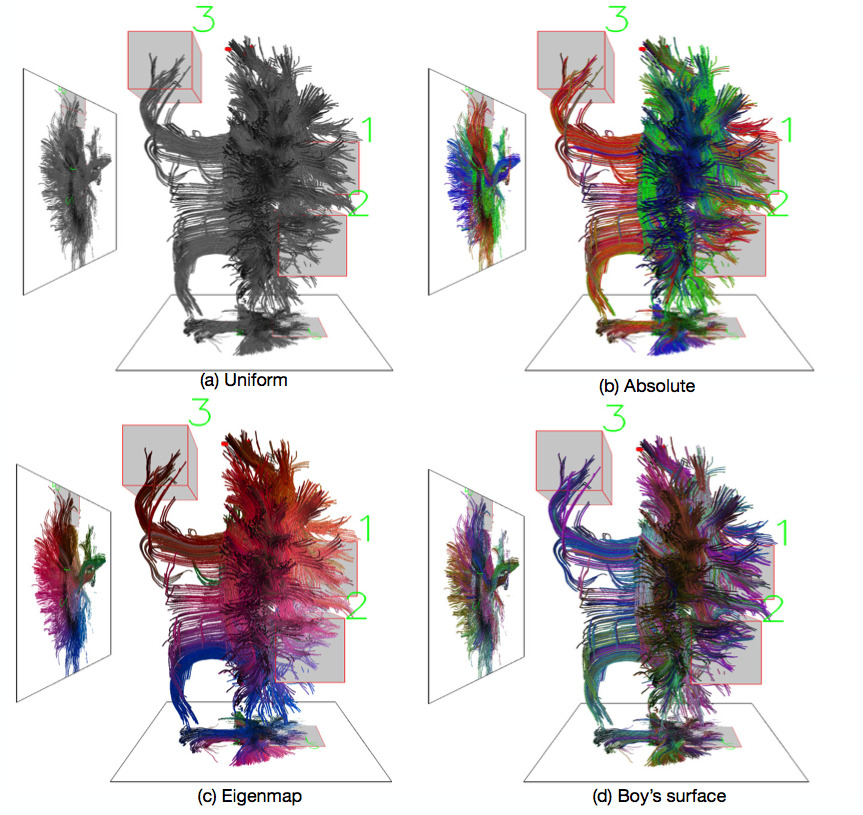}%
\caption{Task 2's Four Orientation Colormaps}
\label{fig:task2maps}
\end{figure}

The four spherical colormaps shown in Figure~\ref{fig:task2maps} are used in task 2 (ensemble tract tracing). 

\textit{Baseline uniform} is used as a control condition.

\textit{Absolute RGB color-triples}
uses Pajevic's approach~\cite{pajevic1999color} in which the three different orientations (left-right, dorsal- ventral, anterior-posterior) are represented as red (R), green (G), and blue (B). Each tract uses a constant color indicating its global orientation. 

\textit{Eigenmap embedding} 
implements the  method of Brun et al.~\cite{brun2003coloring}. It assigns colors to tracts based on the similarities among tracts. 
The tracts becomes points  in the embedded low-dimensional space~\cite{belkin2003laplacian} and 
the similarity of tracts is measured using the closeness of these points and a similarity matrix.  
The 3D coordinates of the points are normalized to fit into the displayable range of the L*A*B* color space and the corresponding colors are used for the tracts.

The \textit{Boy's surface embedding} 
implements the method of Demiralp et al.~\cite{demiralp2009coloring}, a one-to-one mapping between an orientation and a location in a color space based on a Boy's surface immersion in the color space. The embedding is also \textit{angular uniform}, i.e., the larger the difference in tract orientations, the larger the perceptual difference in their colors.

\subsection{Diffusion MRI Datasets}

\begin{figure}[!tp]
\centering
\includegraphics[width=0.7\linewidth]{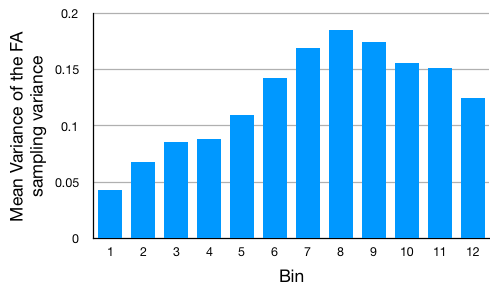}%
\caption{Domain-Specific Data Attributes: The spreads (variances) of all AverageFA data in each of the 12 bins in our random samples, are smaller when AverageFA is in the lower bins and become most spread (with larger variances) when the bins ids $\in$ $[7, 9]$. From bin 1 to bin 12, the average FAs are 1: [0.25, 0.3), 2: [0.3, 0.35), 3: [0.35, 0.4), ......, 12: [0.8, 0.85] respectively.}
\label{fig:dataSpread}
\end{figure}

For task type 1, the average FA values are in the range [0.25, 0.85]. We evenly divided this range 
into 12 bins and the step size was 0.05.
We randomly sampled within the four brain regions (here corpus callosum (CC), cortical spinal tracts (CST), inferior frontal occipital fasciculus (IFO), and inferior longitudinal occipitotemporal fasciculus (ILF)) by randomly placing boxes in these regions.
We then take an equal number of samples in each bin from these samples.  
Fig.~\ref{fig:dataSpread} shows the variance of the data in these 12 bins. We see that the lower and higher mean FA would have narrower spread (smaller variance) than those in the middle; this is the unique domain-specific data attribute.

Because ensemble mean is affected by 
variance{~\cite{maule2015effects}}, one way to conduct a study is to control the variance in each bin and measure the color effectiveness in each bin. 
We did not do this in order to 
retain a high-fidelity representation of tractography features; 
otherwise, we would have to produce artificial data to control the spread in each bin.

For task type 2, tractography data were computed from source DMRI images captured from a normal human brain at resolution $0.9375 mm\times0.9375 mm\times4.52 mm$. Data are also sampled from four major bundles CC, CST, IFO, and ILF.
All tracts are rendered using tubes.

%


\subsection{Experimental Design}

Within-participant design was used for both tasks: i.e., each participant examined all colormaps. 
The independent variable is colormap.  
The dependent variables are completion time, accuracy, and subjective ratings.
For task type 1 of ensemble average with 6 colormaps, each  participant performed 12 instances (in each of the 12 bins) using each of the 6 maps (72 trials). Six instances of data (two CST, two CC, one ILF, and one IFO sample) and the six maps form a Latin square. No data was repetitively used by the same participant. 

For task type 2 of the ensemble set using four colormaps, each participant performed eight instances of each coloring condition with four instances of each of the four bundles (32 trials). Again, datasets were not reused by the same participant. We ordered the four bundles and the four colormaps by a 
$4\times4$ Latin square. 
The order of the trials for each colormap was randomized.

Each participant performed ${72 + 32 = 104}$ sub-tasks. 

\subsection{Participants, Apparatus, and Environment}

A total of 24 participants (17 male and 7 female) took part in the study: two  medical  professionals,  seven  computer science students,  and 15 students from other disciplines (mechanical engineering, math, and global studies). Their average age was 27.8 years  with  standard  deviation 4.0. All participants had normal or corrected-to-normal vision and normal color vision tested using Ishihara Color Test. 

The program runs on a Linux desktop with a $27"$ monitor (BenQ GTG XL 2720Z, resolution $1920 \times 1080$). Gamma was adjusted daily to ensure uniform perceived brightness: the gamma value used for the display was 2.2. 

The lighting used fixed-pipeline 
OpenGL rendering with per-vertex lighting and Gouraud shading. We used a traditional three-point lighting scheme. Key and fill lights were placed in relation to a preset camera with 35mm focal length and the key light is at the top left of the scene, the location assumed by most human observers. Lighting placement and intensity are chosen to generate images with contrast and lighting properties appropriate for the data and human assumptions. For example, the key and fill lights are elevated and slightly to the left and right of the observer. All lights were white.
The screen background color was white.

\subsection{Procedure}

Participants were tested for normal vision and passed the Ishihara Color Vision test. They 
received general information about brain structure and about DMRI techniques and their medical uses. The training session, which lasted about 15 minutes, ensured that the participants understood the coloring and tasks. 


Task completion time was recorded from the time when the visualization was shown on the screen to the time when the final answer button was clicked. Participants were told to be as accurate and as fast as possible, and that accuracy was more important than time.
They were also told to rotate the data to better interpret the structures.
They had to finish a task in order to go to the next one. No time limit was set on each task. 
They could take a break at any time. After finishing all sub-tasks using each colormap, they selected from a 7-point scale (1 (worst) to 7 (best)) on the computer screen to rate the map they just used. Finally, participants were interviewed for their comments. Participants took about an hour on average to finish this study and received monetary compensation. No fatigue was reported.

We conducted three pilot studies comparing performance with a total of 50 participants (including 3 brain scientists) to refine our experimental procedure. These pilot study participants were not used in the formal study. We recruited brain scientists to collect some domain-specific comments related to brain sciences on the color encoding methods. The main difference between expert and novice groups, as observed in our previous study and the pilot studies, was that experts took longer to complete task because they were more interested in examining the data. Our pilot studies revealed no significant difference in task completion time and accuracy between  medical school students and other college students without medical backgrounds.

\section{Results}

\begin{table}[!t]
\caption{Main Effects of Colormap on Accuracy and Task Completion Time and Effect Size. Here C stands for color and P for participant. 
The \textit{large} effect sizes are in \textbf{bold} and the medium ones in \textit{italic}.}
\label{table:exp1:stat} 
\begin{tabular}{| l | l | l | l |}
\hline
 
 Average & C on error & F(5, 1728)=0.98, p=0.43                  & d=0.16\\
               & C on time  & \textbf{F(5, 1728)=6.23, p$<$0.0001} & \textit{d=0.31}\\
               & P on error &  \textbf{F(23, 1728)=2.77, p$<$0.0001}  & \textbf{d=0.71} \\
               & P on time   & \textbf{F(23, 1728)=50.24, p$<$0.0001} & \textbf{d=3.72}\\
\hline
Orientation    & C on error & \textbf{${\chi}^2$ (3, 768)=13.94, p=0.0030} & V=0.13\\
                & C on time & ${\chi}^2$ (3, 768)=0.67, p=0.57 & d=0.13\\
                & P on error & ${\chi}^2$ (23, 768)=23.47, p=0.43 & V=0.17 \\
                & P on time  & \textbf{${\chi}^2$ (23, 768)=4.35, p$<$0.0001} & \textbf{d=1.56}\\
\hline
\end{tabular}
\end{table}

\begin{figure}[t!]
\centering
\subfigure[Absolute Error]{\includegraphics[width=0.78\linewidth]{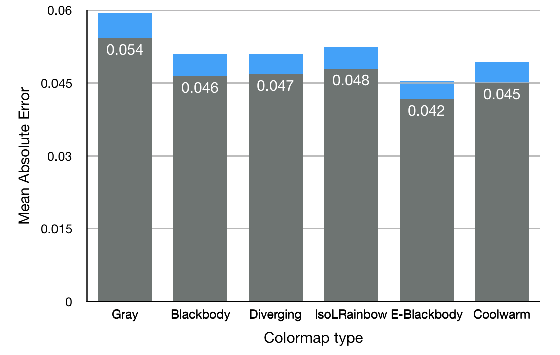}\label{fig:t1error}}

\subfigure[Absolute Task Completion Time]{
 \includegraphics[width=0.75\linewidth]{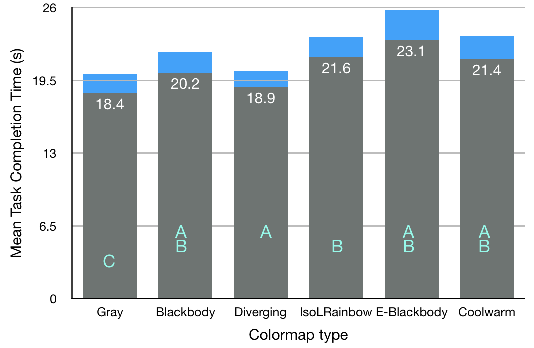}\label{fig:t1time}}

\label{fig:t1results}
\caption{Task 1: Mean Absolute Error and Task Completion Time. The \textcolor{blue}{blue} bars show $95\%$ confidence intervals. 
(A). $Absolute\; error$ = $\left| participant's\; answer - ground\; truth \right|$. (B). Colormaps labeled with the same \textcolor{cyan}{cyan} letter belong to the same group in the post-hoc analysis.}
\end{figure}

We collected 2496 data points with 24 participants for the two ensemble tasks, or 1728 and 768 for  the \textit{ensemble average} and \textit{ensemble set} tasks accordingly.  
To summarize, the first hypothesis (H1 on rainbow) is partially supported.  H2 on multihue, H3 on gray, and H5 on colorfulness are supported. We find 
no evidence to support H4 on resolution.

\begin{figure}[!t]
\centering
\includegraphics[width=0.85\linewidth]{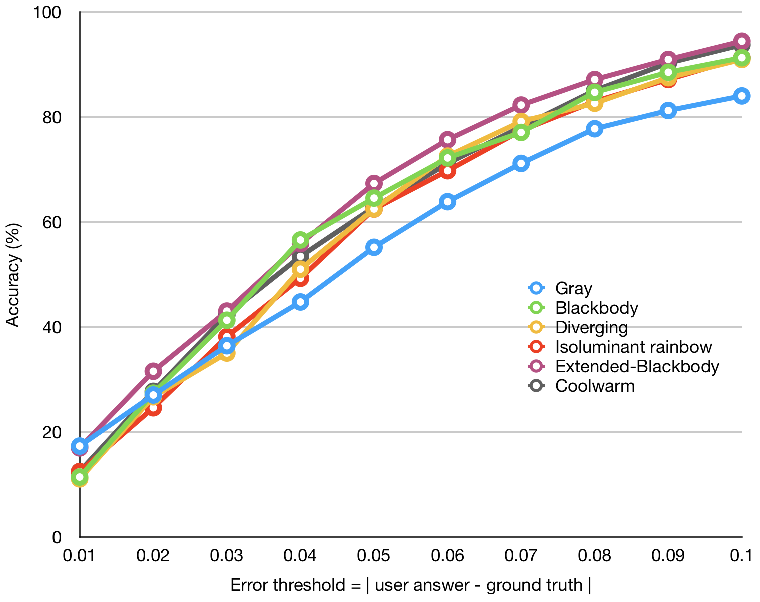}%
\caption{Colormap Accuracy}
\label{fig:t1accuracy}
\end{figure}

\subsection{Overview of Analysis Approaches and Summary Statistics}

Results were analyzed by tasks; Table~\ref{table:exp1:stat} shows the statistical analysis of accuracy and task completion time measured
using the following statistical approaches. For both tasks, we examine the main effect of colormap on error and task completion time using the SAS GLM procedure. A post-hoc analysis using the Tukey Studentized Range test (HSD) is performed when we observe significant main effects. 

Task 1 performance is analyzed using several methods.
Task completion time is converted to $log_{10}$-based to obtain a close-to-normal distribution. We compute \textit{error} by the distance from the participants' answers to the ground truth and use the formula $error$ =  $\log _2 \left| {participant's\; answer} - {ground\; truth} \right| + 8 $, following Cleveland and McGill~\cite{cleveland1984graphical}. 
We explore the accuracy of these ensemble colormaps using two additional measurements. 

\begin{itemize}

\item Accuracy. 
Accuracy is percentage of correct answers. We threshold the error to measure whether an answer is correct.
We used 
$\delta = \left|participant's \;answer - ground\; truth\right|$ and 
threshold $\delta$ to 0.01-0.04 with step size 0.01.
An answer is considered correct when it falls in $\delta$.

\item Directional Bias. We compute whether or not the colormaps bias observers towards values larger or smaller than ground truth.

\end{itemize}

The accuracy data in Task 2 are binary and are analyzed
using logistic regression and reported using the $p$ value from the 
Wald $\chi ^2$  test. When the $p$ value is less than $0.05$, variable  levels  with  $95\%$ confidence  interval  of  pairwise
difference of odds ratios not overlapping are considered significantly different. The $\chi ^2$  test with the ``$freq$'' procedure is used to examine whether or not there is a significant correlation between the main effect (the colormap or participant) and accuracy.

We measure effect sizes using Cohen's $d$  for time and task type I error and Cramer's $V$ for correctness to understand the practical significance~\cite{cohen1988statistical}. We used Cohen’s benchmarks for 
``small''(0.07-0.21), ``medium'' (0.21-0.35), and ``large'' ($>0.35$) effects.

\begin{figure}[!t]
\centering
\centering
\includegraphics[width=0.98\linewidth]{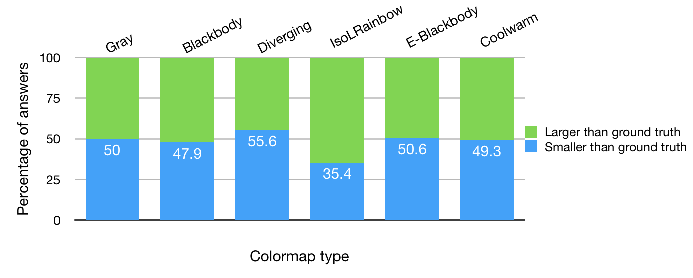}%
\caption{Directional Biases by Colormap.  More participants tend to overshoot (report larger than the ground truth) when using 
isoluminant rainbow. Using the diverging colormap, more participants underestimated the ensemble average. 
Gray, extended-blackbody, and coolwarm had the minimum directional biases.}
\label{fig:t1bias}
\end{figure}

\begin{figure}[!tp]
\centering
\centering
\includegraphics[width=\linewidth]{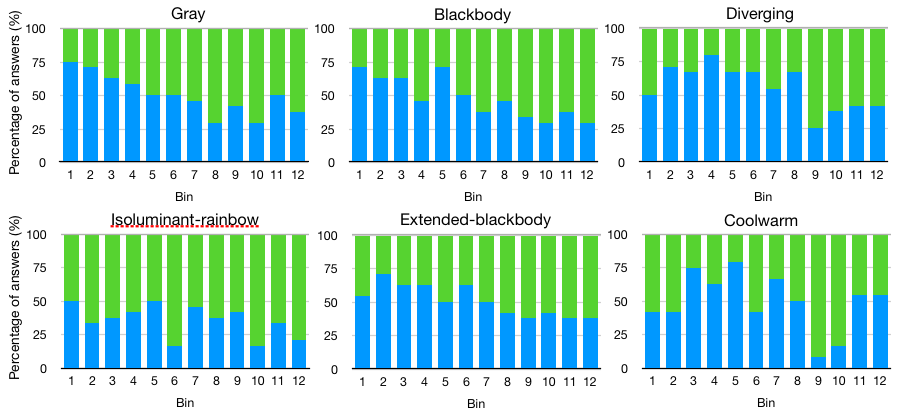}%
\caption{
Directional Biases by Colormap and Bin.
More than $50\%$ larger-than-ground-truth answers appeared in \textit{all} 12 bins for isoluminant rainbow.
}
\label{fig:biasTowardsHigher}
\end{figure}

\subsection{Task 1 Ensemble Average Results}

For task type 1, ensemble average, 
colormaps was not a significant main effect on error (Table~\ref{table:exp1:stat} and Fig.~\ref{fig:t1error}). A general trend was that extended blackbody had the least error and gray had the most.

Colormap and participant are significant main effects on time.
(Table~\ref{table:exp1:stat} and Figure~\ref{fig:t1time}). 
The post-hoc analysis suggests three Tukey groups: (gray), (blackbody, isoluminant-rainbow, extended-blackbody, and coolwarm), and (blackbody, diverging, extended-blackbody, and coolwarm). The extended-blackbody and coolwarm maps led to the longest task completion time and the gray, though efficient, had the highest error.

\subsection{Task 1 Color Sensitivity and Directional Bias}

We compute the colormap sensitivity by measuring the percentage of correct answers or accuracy (Fig.~\ref{fig:t1accuracy}). 
We first compute the mean absolute error.
Fig.~\ref{fig:t1accuracy} showed that
gray had on average the lowest accuracy among all colormaps. 
 

Directional bias measures if observers consistently choose larger or smaller values than the ground truth using a colormap. We found that 
more answers using isoluminant rainbow were biased towards higher values, while the diverging color slightly towards lower answers (Fig.~\ref{fig:t1bias}). All other colormaps of blackbody, extended-blackbody, and coolwarm showed about even distributions between higher and lower participants' answers.

We further analyzed the bias distribution in the 12 bins (Fig.~\ref{fig:biasTowardsHigher}).
We found that more than $50\%$ of the answers overshoot (selected larger than ground-truth) when
using isolumiant-rainbow in \textit{all} bins.
Correlations between the data variance and colormap absolute error show that these two variables are statistically significantly correlated for all other maps except the isoluminant-rainbow.  This result may indicate that the ensemble behaviors of isoluminant-rainbow might
not be as predicable, despite its accuracy for ensemble average is comparable to other colormaps.


\subsection{Task 2 Ensemble Spherical Colormap Results}

\begin{figure}[!tp]
\centering
\subfigure[Absolute Accuracy]{
\includegraphics[width=0.6\linewidth]{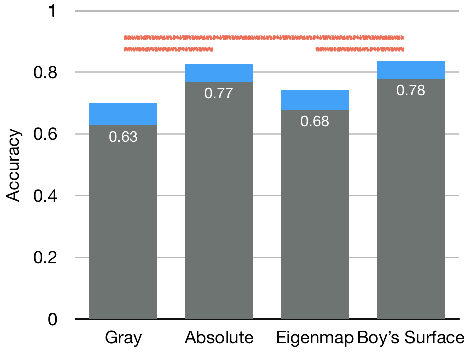}%
\label{fig:t2accuracy}}

\subfigure[Absolute Task Completion Time]{
 \includegraphics[width=0.58\linewidth]{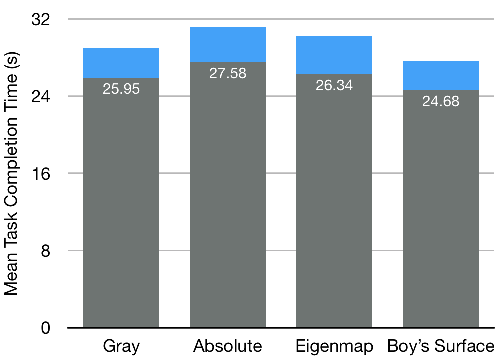}%
\label{fig:t2time}}

\label{fig:t2results}
\caption{Task 2: Mean Time and Accuracy. The color schemes connected by the \textcolor{orange}{orange} line are significantly different.}
\end{figure}

The second row in  Table~\ref{table:exp1:stat} shows the  statistical  results.
Fig.~\ref{fig:t2accuracy}  shows mean accuracy (percentage correct answers) and time and $95\%$ confidence intervals from the mean. Colormap had a significant main effect on  accuracy  but not on task completion time. 
H4 is not supported.
The Boy's surface embedding and the absolute embedding lead to most accurate answers for following tracts, followed by eigenmap. Boy's surface also shortened task completion time. This task does not require participants to utilize symmetry.
The Boy's surface method was more accurate and also fast (Figs.{~\ref{fig:t2accuracy}} and{~\ref{fig:t2time}}). It is also noticeable that the Boy's surface and absolute maps improved accuracy by  $15\%$ and $14\%$ respectively compared to the baseline gray.
Our results support the last hypothesis (colorfulness hypothesis) since all maps with colors increase accuracy over the gray baseline.

\begin{figure}[!tp]
\centering
\includegraphics[width=0.98\linewidth]{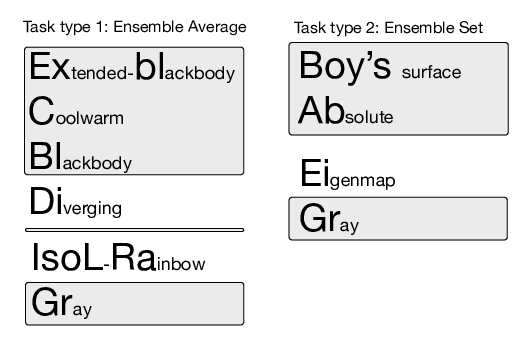}%
\caption{Ensemble Ranking of Visualization Methods.}
\label{fig:ranking}
\end{figure}

\subsection{Subjective Ratings and Comments}
Participants' ratings and comments provide useful insights into how the  usefulness  of  the  colormaps was perceived. 
Participants' subjective rating of the usefulness of these colormaps, from high to low are: task1: coolwarm (5), extended-blackbody   (4.96),   blackbody   (4.96),   diverging (4.75), isoluminant-rainbow  (4.4),  and  grayscale  (3.7);  task2: absolute (5.3), eigenmap (5.3), Boy's surface (4.8), and uniform uniform-gray (2). 
Grayscale in task 1 and uniform gray with no coloring was rated least useful for both tasks.

%

The interviews revealed that  those  who  liked  the  \textit{absolute} method found it the simplest to understand and  easiest for following the tracts because of its symmetry; in addition, the  less  chaotic  color  changes helped them  recognize the orientations better.  Those who disliked the absolute method thought that tracts looked  too similar  to differentiate, show- ing  the  tradeoffs between similarity and  resolution. Most participants were  relatively neutral on  the \textit{Boy's surface},  considering it similar  to  the  \textit{eigenmap} method in terms of hue uses (spatial resolution) despite including more hues   than   that  method. Participants commented that ``\textit{it (Boy's surface) was useful to have some different hues, but too many hues made the visualization less intuitive}'', while  others stated that  the  ``\textit{right  amount  of hues of eigenmap provided enough discriminations between values without overloading one?s perception capability.}''


\begin{figure*}[!th]
\centering
\includegraphics[width=0.99\linewidth]{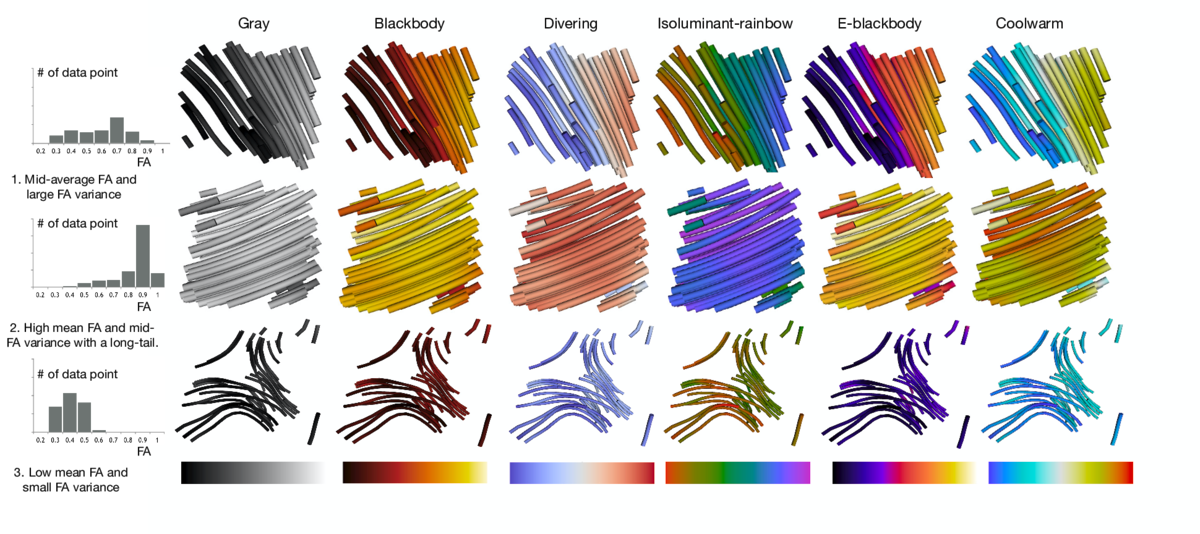}%
\caption{Example Dataset Distribution and Their Colormaps: top: high-variance; middle: higher mean FA and narrow long-tail; bottom: low mean FA and narrow variance.}
\label{fig:distribution}
\end{figure*}

\section{Discussion}

This section discusses our results. Fig.~\ref{fig:ranking} shows our 
recommendations for choosing colormaps for the two ensemble tasks studied here.

\begin{figure*}[htp]
\centering
\includegraphics[width=\linewidth]{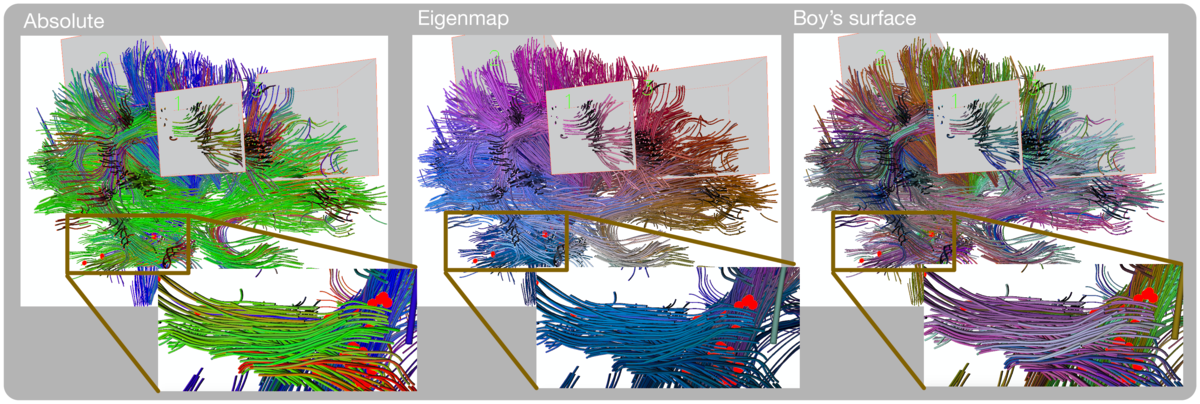}%
\caption{An example from the empirical study for which all participants got correct answers using absolute and Boy's surface but only half the participants got correct answers with Eigenmap. Red dots in the subfigures are sources. Eigenmap tends to show similar colors in cases in which the other two methods produce visually distinguishable ones.
}
\label{fig:eigenBadcase}
\end{figure*}

\subsection{Isoluminant Rainbow Does Not Decrease the Mean Accuracy, but Introduce Biases}

Our first hypothesis
is only partially supported.
The most interesting result may be that the isoluminant rainbow does 
\textit{not} introduce greater error on average for task 1 (Fig.~\ref{fig:t1accuracy}).  
This efficiency  result  may  agree with  those  in vision  science because humans can average hues   
because humans can average hues{~\cite{maule2014getting}}. However,  none of the 
vision science studies to our knowledge drills down to the empirical study results to examine whether or not participants would be biased towards
higher or lower than ground truth. The fact that  isoluminant rainbow introduces higher overshooting needs  to be further studied, perhaps by  explicitly controlling  the  variance in  data for us to learn  the colormap behaviors. 
Rainbow colors are known to be poor for univariate encoding due to the lack of uniformity and ordering and because they produce
artificial boundaries in data. 
We could conclude from our study that ensemble color processing differs from univariate colormap representations.

We do not recommend this isoluminant-rainbow map for ensemble average tasks. 
Instead, we propose to further explore
\textit{how} and \textit{why} multihue works for limited capacity ensemble processing. 
This is mainly because the biases in isoluminant rainbow are consistent independent of the variances in data (Fig.{~\ref{fig:dataSpread}} and Fig.{~\ref{fig:biasTowardsHigher}}).
The rainbow map certainly uses a set of semantically meaningful colors that would ease human understanding and our brain scientist collaborators particularly love rainbows; 
however, rainbow maps may still violate Trumbo's color design heuristics that \textit{``the basic information should be displayed in a clear and logical fashion so that it may be decoded with precision and without continual references to the key (labeled scheme)''} 
and \textit{``if small neighboring regions produce illusion of color over larger map areas, these illusions should not give misleading information''}~\cite{trumbo1981theory}.

\subsection{Multihue Maps Improve Ensemble Accuracy in General}

Our second hypothesis about multihue efficiency is supported.
We ran a statistical analysis to examine whether or not hue or luminance affect error or task completion time. We found that hue had a significant main effect on time ($F(2, 1728) = 4.99$, $p = 0.0069$). The post-hoc analysis showed that colormaps with multihue led to statistically significantly longer task completion time than single-hue (gray) colormaps. 

The multihue extended-blackbody and the coolwarm colormaps had the lowest absolute error, with slightly longer task completion time. 
This accuracy result of extended-blackbody agrees with 2D study results as well, though we did not observe significant differences. There may be at least two reasons for the benefits. First, one might think these two colormaps had the largest arc-length and thus yielded slightly better results than other maps. 
The other, perhaps primary reason for the benefits is that the multihue lets participants quickly determine the target-region first before formulating their answers, and this two-stage viewing could also explain why rainbows also take longer to execute. Visual inspection of colormaps applied to empirical data in three different FA distributions (Fig.~\ref{fig:distribution}) shows the FA variances when the mean is around the middle (top row), in the higher (middle row) or lower (bottom row) end.  
We may observe that the colormaps in the last three columns with many hues may help viewers quickly locate the target regions on the colormap into which the answers fall. 

\subsection{That Many Colormaps Work Well Also Shows the Power of Human Visual Systems in Judging Ensemble Averages}

We did not observe differences in accuracy among colormaps when measuring the distance of participants' answers from the ground truth. This result suggests the power of visual ensembles for quantitative estimates.

Balancing all considerations of efficiency, error, and correctness, and bias in these colormaps, we rank them in the order shown in Fig.~\ref{fig:ranking} task 1, where extended-blackbody, coolwarm, and blackbody seem to work well. Isoluminant-rainbow and diverging are worth further investigations. Gray is not recommended because of their higher biases.
Though we cannot say whether the poor performance of grayscale was caused by its simultaneous contrast or its sole luminance channel, the result indeed is in agreement with the literature on 2D colorization.

\subsection{Local Contrast and Resolution Together Might Be the Most Decisive Property for Ensemble Direction Tracing}

Our results present an uncanny valley effect where the highest and lowest resolution maps improved outcome compared to the mid-resolution eigenmap orientation map. H4 is not supported.

Overall, our results did not suggest that \textit{resolution} contributes to higher accuracy in 3D space, since both the Boy's surface and absolute
methods reduced errors. The eigenmap had reasonable resolution, as does the Boy's surface colormap, but lowered accuracy.
To understand \textit{when} Boy's surface and absolute succeeded and eigenamps failed. we inspected qualitatively 
by the best and worst examples of participant accuracy when using these colormaps, as shown in Fig.~\ref{fig:eigenBadcase}. 
We see that, while eigenmap provides regional coloring, the adjacent regions have relatively low contrast compared to other two approaches.
These observations may suggest that local contrast is the most decisive property, since a combination of high contrast and spatial resolution, as in the Boy's surface, led to higher accuracy on ensemble tracing. 
Boy's surface generates 
colors that seem to strike the right balance in the spatial resolution and contrast for this spatial structure determination. 
Finally, the data sample varies so no dataset is seen twice by the same participants. 
For the eigenmap, this setting means that the colors for the same tracts in different datasets would change, while the same tube would always be given the same color with the other maps.

We therefore recommend Boy's surface and absolute for coloring DMRI ensemble set, as shown in Fig{~\ref{fig:ranking}}.

\subsection{Reuse of Our Results to Other Ensemble Representations}

We sought to further our understanding of the  ensemble data  processing to  generate concrete  implications for  visual  analysis of brain  DMRI tractography datasets. 
In general, both tasks suggest that high-contrast localized colormaps may have helped both ensemble average and 
tract discrimination.
Reuse  of our  results in other  domains would have  to take into account domain specificities of data, task, and user. 
Several areas could benefit from our work, such as weather forecasting{~\cite{sanyal2010noodles}}, hurricane track prediction{~\cite{cox2013visualizing}}, and motion or movement trajectories{~\cite{chen2009visual}}{~\cite{andrienko2016leveraging}},
because  direct trajectory depiction has been informative.
The most suitable reuse would be when the 
datasets have  relatively low  variance, so that colormaps can be localized to a smaller regions on a colormap for 
scalar data visualizations.
Similarly, the spherical orientation colormap for line field visualizations might also be domain-dependent. In our case, the tracts are
following three major orientations. We also did not consider other 
tract shapes. 
Considering appropriate distance measures is needed for maximal performance.

\subsection{Participants' Experiences}

Participants in this study have different backgrounds, and an ideal condition might be to use only brain scientists, clinicians, or medical school students. One major reason for the background differences was that we had access to only a few brain scientists. We used as many as possible in the study because we wanted to collect their comments related to the brain science domain.

Also, we followed Munzner's approach~\cite{munzner2009nested}  of abstracting tasks into a level suitable for empirical study. In other words, these tasks could be performed by a trained participant. 
This may explain why we did not observe differences in task completion time and accuracy between students with and without medical backgrounds.
Several user studies in flow visualization have used non-domain experts, suggesting that non-domain-expert is a viable option in empirical studies~\cite{liu20122d}. 

\subsection{Using Ensemble for Visualization Design}

It is intuitive to think  that  hue,  due  to its categorical
effect  (e.g. yellow   or  red),  would  interfere with  the ensemble coloring,  thus  making  representing 
a multihue average difficult. However, this turns out not to be the case.
In vision science, ensemble is believed to be used by the human visual system to address our severely limited visual working memory. We can quickly derive patterns that guide our attention towards the most useful information.
Scientific data is often highly structured and may carry redundant structures. When there is redundancy, it is possible to 
sample and filter to produce optimal views.
For example,  a handful of past visualization work has shown that implicit or explicit representation of sets of objects as groups or ensembles can guide observers' attention to process only the most relevant incoming information (e.g., explicit depiction of a group of objects in clusters~\cite{peng2012mesh}, grouping interfaces to augment exploration workflows~\cite{li2011visbubbles}~\cite{ragan2016characterizing} or using spatial patterns to form texture pattern to guide observers' behavior~\cite{zhao2017bivariate}). 
We believe there will be an opportunity to create a compressed and efficient ensemble representation of information, such as ensemble overviews, to guide visual attention to the areas more relevant to the targets.

\subsection{Limitations and Future Work}

The aim of this paper was to investigate the effect of coloring in practice on two spatial ensemble visualization tasks, average and set orientation. 
Our study is only a first step towards understanding ensemble tasks in visualizations. 
Although this study can suggest \textit{what} colormap to choose for ensemble representation, we may need to build computational models or isolate
factors (e.g., hue and luminance for task 1 and resolution and uniqueness for task 2) to explain \textit{how} these colormaps are used by our visual system. 
Effectiveness of these coloring approaches needs to be studied further when tasks are related to other discrimination and detection tasks, in which quantitative differences among data are to be reported.

Our study would suggest further work.
Since multihue colormaps in general improved ensemble average accuracy, one could run studies to systematically control the mean and variance of the ensemble datasets to model the ensemble performance.
Viewers make a two-alternative forced-choice judgment about which visualization
method contains the larger average value. Sensitivities are measured based on the differences between the values. 
A psychometric function fitted to the data reveals sensitivity to the discriminative threshold to measure accuracy. 
Using this method, we could answer questions about \textit{why} and \textit{when} multihue average will be effective and 
how variance influences the effectiveness and efficiency.

\section{Conclusion}

This study is the first (to  our knowledge) to compare different color ensemble encodings for 3D DMRI tractography visualizations. 
Results from the study provide the following insights for choosing 3D tube coloring ensembles.

\begin{itemize}
\item 
The most interesting result was that the isoluminant-rainbow performed reasonably well, though it did lead to more reporting bias towards higher than ground truth values than other colormaps.

\item 
Extended-blackbody, coolwarm, and blackbody are  reasonably accurate for ensemble average in 3D. Our analysis showed that hue had much larger influence on error than luminance.

\item
Our study on the ensemble set orientation discrimination supports the proposition that having some colors is significantly better than no color at all.

\item 
Colormaps with better orientation contrast (e.g., the Boy's surface and the absolute approach) are most desirable for ensemble set orientation discrimination tasks such as tract tracing.

\end{itemize}

\appendices

\begin{figure*}[!t]
\centering
\includegraphics[width=0.98\linewidth]{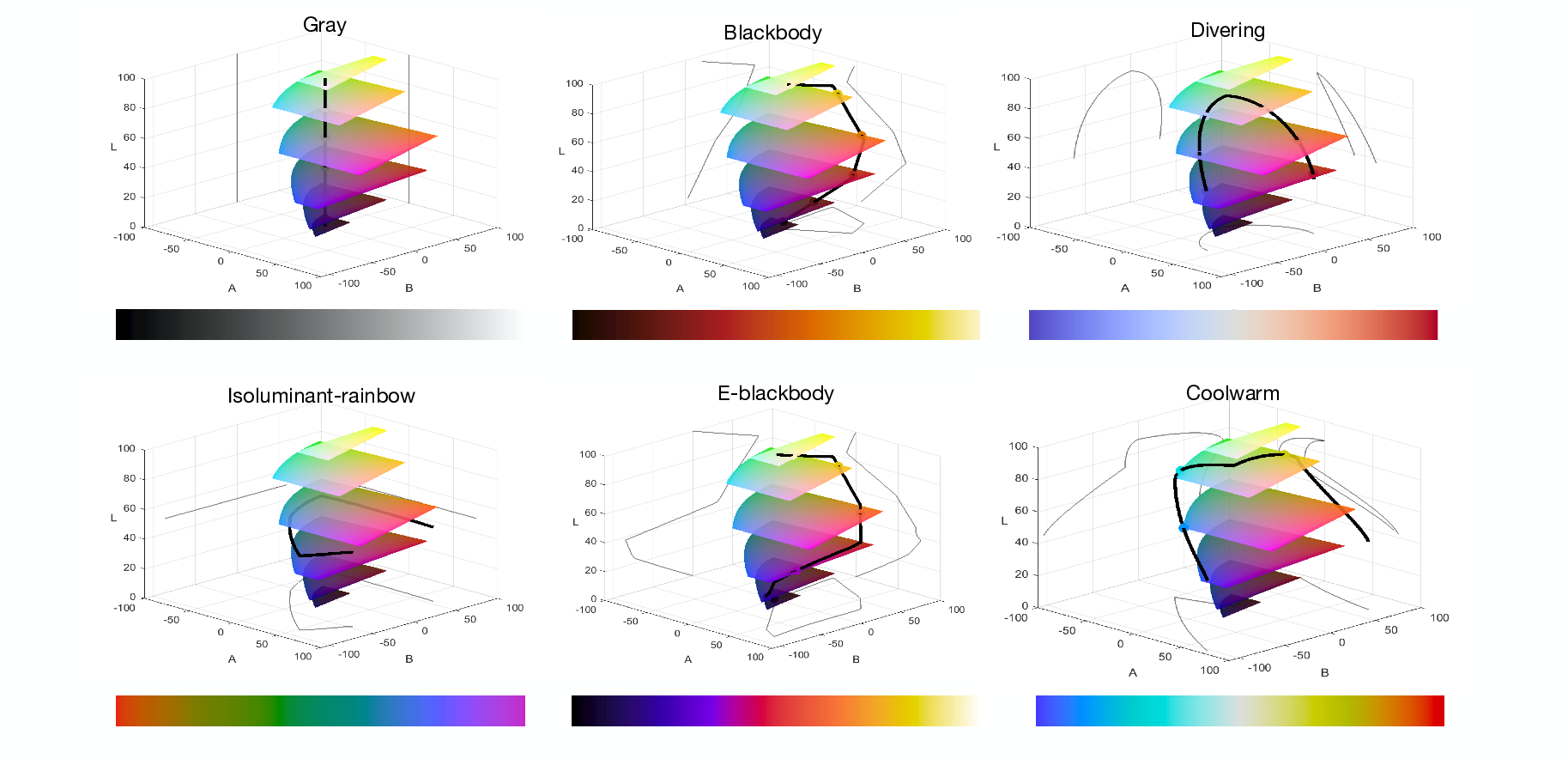}%
\caption{Colormap Profile for Showing Scalars in Task 1 (AverageFA tasks) in the L*A*B* color space. L-planes from bottom to top are L=5, 20, 40, 60, 80, and 95.}
\label{fig:colorSpace}
\end{figure*}

\begin{figure*}[!t]
\centering
\centering
\includegraphics[width=0.98\linewidth]{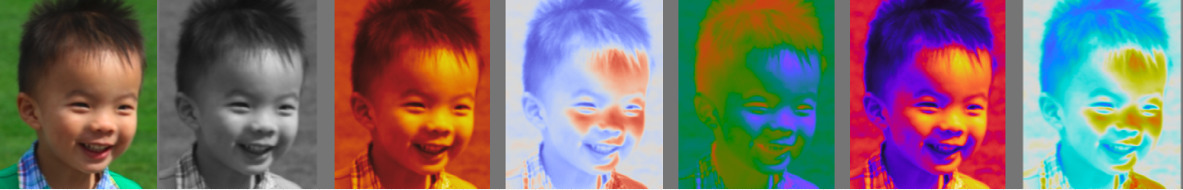}%
\caption{Using Faces to Examine the Luminance Profile of Colormaps (from left to right): original image, gray, blackbody, diverging, isoluminant-rainbow, extended-blackbody, and coolwarm colormaps.}
\label{fig:face}
\end{figure*}

\section{The Univariate Colormaps in the L*A*B* Color Space}

Fig.~\ref{fig:colorSpace} shows the scalar colormaps in the L*A*B* color space. 
The curve in each figure shows the trajectory of color maps and their three projects in the L*A*B* color space. 
All color interpolation is performed using linear interpolation in this space.




We used the Rogowitz-Kalvin{~\cite{rogowitz2001blair}}
and Kindlmann-Reinhard-Creem approaches{~\cite{kindlmann2002face}} to help visually inspect colormaps to test their luminance profile. 
This method utilizes our sensitivity to luminance variations in human faces to select colormaps. 
Fig.{~\ref{fig:face}} shows samples of faces generated by these six colormaps with our online tool.The faces with isoluminance-rainbow and diverging colormaps are
less recognizable than all others. The rainbow and coolwarm colormaps help distinguish different values: one can clearly see red (high) values around the nose and under the eyes.


\section{Coloring Tool Website}

\begin{figure}[!tp]
\centering
\includegraphics[width=\linewidth]{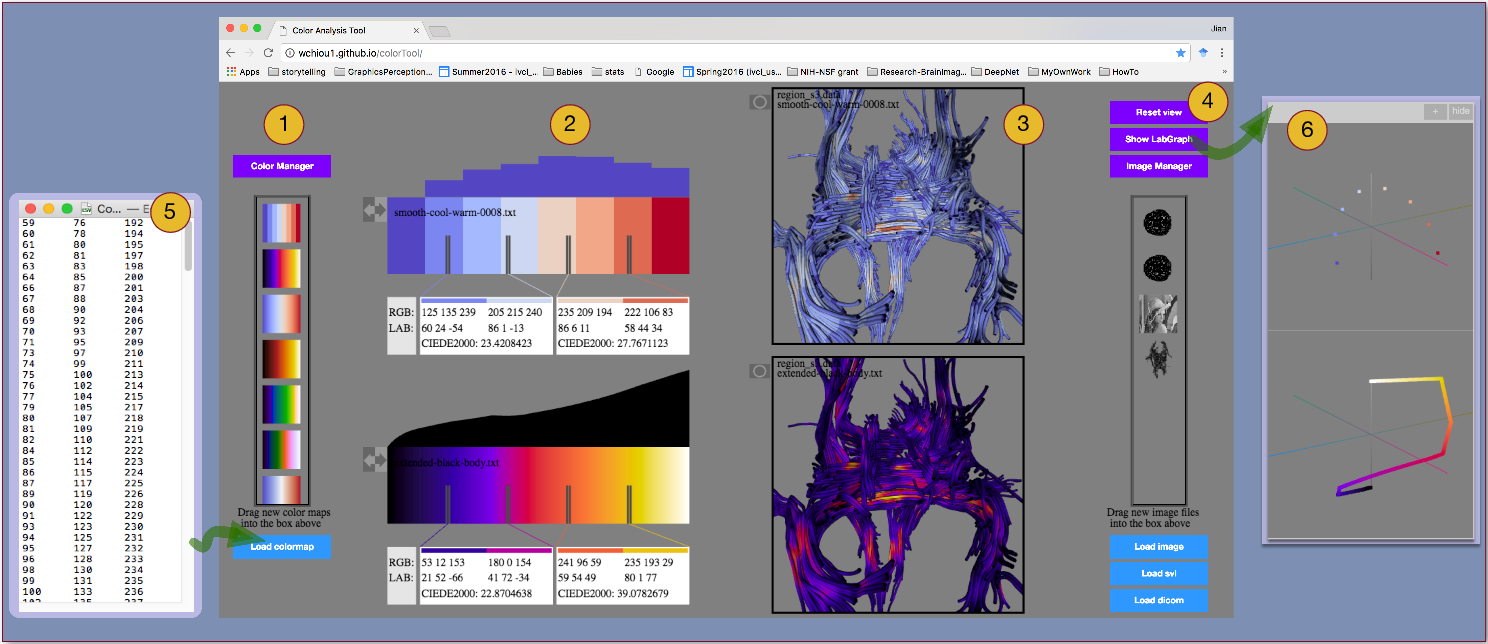}%
\caption{Exploratory Color Comparison Tool.}
\label{fig:tool}
\end{figure}

Our own tool (Fig.{~\ref{fig:tool}}) is hosted at {http://wchiou1.github.io/colorTool/({Fig.~\ref{fig:tool}})}. During the evaluation process, we found that using a coloring tool to quickly provide side-by-side comparison made our discussion with the medical doctors very effective and efficient.  The direct manipulation interface lets users directly drag and drop plain-text colormaps. It can display both 2D image and 3D geometry examples.

\ifCLASSOPTIONcompsoc
  \section*{Acknowledgments}
\else
  \section*{Acknowledgment}
\fi

The authors would like to thank Drs. Peter Kochunov, L. Elliot Hong, and Neda Jahanshad for their discussions on  color uses in brain science, Dr. Bernice Rogowitz for a discussion on  colormap uses in real-world applications, Dr. Jeremy Wolfe for his thorough review and comments on this manuscript, and the anonymous reviewers for their constructive comments.
The authors also thank the participants at University of Maryland, Baltimore Country, University of Maryland Medical School, and Veterans Affairs Medical Center of Providence, RI for their time and effort. We thank Katrina Avery for her editorial support. 
This work was  supported  in  part by NSF IIS-1302755, CNS-1531491, DBI-1260795, IIS-1018769, and DUE-0817106 and by NIST MSE-70NANB13H181. Any opinions, findings, and conclusions or recommendations expressed in this material are those of the authors and do not necessarily reflect the views of National Institute of Standards and Technology (NIST) or the National Science Foundation (NSF.)

Jian Chen is the corresponding author.

\ifCLASSOPTIONcaptionsoff
  \newpage
\fi

\bibliographystyle{IEEEtran}
\bibliography{main}

\begin{thebibliography}{10}
\providecommand{\url}[1]{#1}
\csname url@samestyle\endcsname
\providecommand{\newblock}{\relax}
\providecommand{\bibinfo}[2]{#2}
\providecommand{\BIBentrySTDinterwordspacing}{\spaceskip=0pt\relax}
\providecommand{\BIBentryALTinterwordstretchfactor}{4}
\providecommand{\BIBentryALTinterwordspacing}{\spaceskip=\fontdimen2\font plus
\BIBentryALTinterwordstretchfactor\fontdimen3\font minus
  \fontdimen4\font\relax}
\providecommand{\BIBforeignlanguage}[2]{{%
\expandafter\ifx\csname l@#1\endcsname\relax
\typeout{** WARNING: IEEEtran.bst: No hyphenation pattern has been}%
\typeout{** loaded for the language `#1'. Using the pattern for}%
\typeout{** the default language instead.}%
\else
\language=\csname l@#1\endcsname
\fi
#2}}
\providecommand{\BIBdecl}{\relax}
\BIBdecl

\bibitem{phadke2012exploring}
M.~N. Phadke, L.~Pinto, F.~Alabi, J.~Harter, R.~M. Taylor~II, X.~Wu,
  H.~Petersen, S.~A. Bass, and C.~G. Healey, ``Exploring ensemble
  visualization,'' in \emph{Proceedings of {SPIE}}, vol. 8294, no. 82940B (12
  pages), 2012.

\bibitem{leib2016fast}
A.~Y. Leib, A.~Kosovicheva, and D.~Whitney, ``Fast ensemble representations for
  abstract visual impressions,'' \emph{Nature Communications}, vol. 7 (article
  number 13186), 2016.

\bibitem{chetverikov2017representing}
A.~Chetverikov, G.~Campana, and {\'A}.~Kristj{\'a}nsson, ``Representing color
  ensembles,'' \emph{Psychological Science}, vol.~28, no.~10, pp. 1510--1517,
  2017.

\bibitem{ariely2001seeing}
D.~Ariely, ``Seeing sets: Representation by statistical properties,''
  \emph{Psychological Science}, vol.~12, no.~2, pp. 157--162, 2001.

\bibitem{robitaille2011more}
N.~Robitaille and I.~M. Harris, ``When more is less: Extraction of summary
  statistics benefits from larger sets,'' \emph{Journal of Vision}, vol.~11,
  no.~12, pp. 1--8, 2011.

\bibitem{alvarez2008representation}
G.~A. Alvarez and A.~Oliva, ``The representation of simple ensemble visual
  features outside the focus of attention,'' \emph{Psychological Science},
  vol.~19, no.~4, pp. 392--398, 2008.

\bibitem{williams1984coherent}
D.~W. Williams and R.~Sekuler, ``Coherent global motion percepts from
  stochastic local motions,'' \emph{Vision Research}, vol.~24, no.~1, pp.
  55--62, 1984.

\bibitem{watamaniuk1992human}
S.~N. Watamaniuk and A.~Duchon, ``The human visual system averages speed
  information,'' \emph{Vision Research}, vol.~32, no.~5, pp. 931--941, 1992.

\bibitem{burr2008visual}
D.~Burr and J.~Ross, ``A visual sense of number,'' \emph{Current Biology},
  vol.~18, no.~6, pp. 425--428, 2008.

\bibitem{neumann2013viewers}
M.~F. Neumann, S.~R. Schweinberger, and A.~M. Burton, ``Viewers extract mean
  and individual identity from sets of famous faces,'' \emph{Cognition}, vol.
  128, no.~1, pp. 56--63, 2013.

\bibitem{oliva2006building}
A.~Oliva and A.~Torralba, ``Building the {GIST} of a scene: The role of global
  image features in recognition,'' \emph{Progress in Brain Research}, vol. 155,
  pp. 23--36, 2006.

\bibitem{bauer2009does}
B.~Bauer, ``Does {Stevens}'s power law for brightness extend to perceptual
  brightness averaging?'' \emph{The Psychological Record}, vol.~59, no.~2, p.
  171, 2009.

\bibitem{chua2008diffusion}
T.~C. Chua, W.~Wen, M.~J. Slavin, and P.~S. Sachdev, ``Diffusion tensor imaging
  in mild cognitive impairment and {Alzheimer}'s disease: a review,''
  \emph{Current Opinion in Neurology}, vol.~21, no.~1, pp. 83--92, 2008.

\bibitem{zhou2016survey}
L.~Zhou and C.~D. Hansen, ``A survey of colormaps in visualization,''
  \emph{IEEE Transactions on Visualization and Computer Graphics}, vol.~22,
  no.~8, pp. 2051--2069, 2016.

\bibitem{silva2011using}
S.~Silva, B.~S. Santos, and J.~Madeira, ``Using color in visualization: A
  survey,'' \emph{Computers \& Graphics}, vol.~35, no.~2, pp. 320--333, 2011.

\bibitem{finlayson2001color}
G.~D. Finlayson, S.~D. Hordley, and P.~M. Hubel, ``Color by correlation: A
  simple, unifying framework for color constancy,'' \emph{IEEE Transactions on
  Pattern Analysis and Machine Intelligence}, vol.~23, no.~11, pp. 1209--1221,
  2001.

\bibitem{kim2009modeling}
M.~H. Kim, T.~Weyrich, and J.~Kautz, ``Modeling human color perception under
  extended luminance levels,'' in \emph{ACM SIGGRAPH}, vol.~28, no. 3 (10
  pages), 2009.

\bibitem{moreland2016we}
K.~Moreland, ``Why we use bad color maps and what you can do about it,''
  \emph{Electronic Imaging}, vol. 2016, no.~16, pp. 1--6, 2016.

\bibitem{christen2013colorful}
M.~Christen, D.~A. Vitacco, L.~Huber, J.~Harboe, S.~I. Fabrikant, and
  P.~Brugger, ``Colorful brains: 14years of display practice in functional
  neuroimaging,'' \emph{NeuroImage}, vol.~73, pp. 30--39, 2013.

\bibitem{szafir2016four}
D.~A. Szafir, S.~Haroz, M.~Gleicher, and S.~Franconeri, ``Four types of
  ensemble coding in data visualizations,'' \emph{Journal of Vision}, vol.~16,
  no.~5, pp. 1--19, 2016.

\bibitem{pajevic1999color}
S.~Pajevic, C.~Pierpaoli \emph{et~al.}, ``Color schemes to represent the
  orientation of anisotropic tissues from diffusion tensor data: application to
  white matter fiber tract mapping in the human brain,'' \emph{Magnetic
  Resonance in Medicine}, vol.~42, no.~3, pp. 526--540, 1999.

\bibitem{demiralp2009coloring}
C.~Demiralp, J.~F. Hughes, and D.~H. Laidlaw, ``Coloring {3D} line fields using
  boy's real projective plane immersion,'' \emph{IEEE Transactions on
  Visualization and Computer Graphics}, vol.~15, no.~6, pp. 1457--1464, 2009.

\bibitem{potter2009ensemble}
K.~Potter, A.~Wilson, P.-T. Bremer, D.~Williams, C.~Doutriaux, V.~Pascucci, and
  C.~R. Johnson, ``Ensemble-vis: A framework for the statistical visualization
  of ensemble data,'' in \emph{IEEE International Conference on Data Mining
  Workshops}, 2009, pp. 233--240.

\bibitem{whitney2017ensemble}
D.~Whitney and A.~Y. Leib, ``Ensemble perception,'' \emph{Annual Review of
  Psychology}, vol.~69, no.~12, pp. 1--25, 2017.

\bibitem{alvarez2011representing}
G.~A. Alvarez, ``Representing multiple objects as an ensemble enhances visual
  cognition,'' \emph{Trends in cognitive sciences}, vol.~15, no.~3, pp.
  122--131, 2011.

\bibitem{potter2012quantification}
K.~Potter, P.~Rosen, and C.~Johnson, ``From quantification to visualization: A
  taxonomy of uncertainty visualization approaches,'' \emph{Uncertainty
  Quantification in Scientific Computing}, pp. 226--249, 2012.

\bibitem{sanyal2010noodles}
J.~Sanyal, S.~Zhang, J.~Dyer, A.~Mercer, P.~Amburn, and R.~Moorhead, ``Noodles:
  A tool for visualization of numerical weather model ensemble uncertainty,''
  \emph{IEEE Transactions on Visualization and Computer Graphics}, vol.~16,
  no.~6, pp. 1421--1430, 2010.

\bibitem{laramee2004state}
R.~S. Laramee, H.~Hauser, H.~Doleisch, B.~Vrolijk, F.~H. Post, and D.~Weiskopf,
  ``The state of the art in flow visualization: Dense and texture-based
  techniques,'' in \emph{Computer Graphics Forum}, vol.~23, no.~2, 2004, pp.
  203--221.

\bibitem{max1990area}
N.~Max, P.~Hanrahan, and R.~Crawfis, ``Area and volume coherence for efficient
  visualization of 3d scalar functions,'' \emph{Proceedings of the Workshop on
  Volume visualization}, pp. 27--33, 1990.

\bibitem{forsberg2009comparing}
A.~Forsberg, J.~Chen, and D.~H. Laidlaw, ``Comparing {3D} vector field
  visualization methods: A user study,'' \emph{IEEE Transactions on
  Visualization and Computer Graphics}, vol.~15, no.~6, pp. 1219--1226, 2009.

\bibitem{haberman2012ensemble}
J.~Haberman and D.~Whitney, ``Ensemble perception: Summarizing the scene and
  broadening the limits of visual processing,'' \emph{From Perception to
  Consciousness: Searching with Anne Treisman}, pp. 339--349, 2012.

\bibitem{maule2015effects}
J.~Maule and A.~Franklin, ``Effects of ensemble complexity and perceptual
  similarity on rapid averaging of hue,'' \emph{Journal of vision}, vol.~15,
  no.~4, pp. 6--6, 2015.

\bibitem{haberman2015individual}
J.~Haberman, T.~F. Brady, and G.~A. Alvarez, ``Individual differences in
  ensemble perception reveal multiple, independent levels of ensemble
  representation.'' \emph{Journal of Experimental Psychology: General}, vol.
  144, no.~2, pp. 432--446, 2015.

\bibitem{correll2012comparing}
M.~Correll, D.~Albers, S.~Franconeri, and M.~Gleicher, ``Comparing averages in
  time series data,'' in \emph{Proceedings of ACM SIGCHI}, 2012, pp.
  1095--1104.

\bibitem{cleveland1984graphical}
W.~S. Cleveland and R.~McGill, ``Graphical perception: Theory, experimentation,
  and application to the development of graphical methods,'' \emph{Journal of
  the American statistical Association}, vol.~79, no. 387, pp. 531--554, 1984.

\bibitem{webster2014perceiving}
J.~Webster, P.~Kay, and M.~A. Webster, ``Perceiving the average hue of color
  arrays,'' \emph{JOSA A}, vol.~31, no.~4, pp. A283--A292, 2014.

\bibitem{wright2011effects}
O.~Wright, C.~Biggam, C.~Hough, C.~Kay, and D.~Simmons, ``Effects of stimulus
  range on color categorization,'' \emph{New directions in colour studies.
  Amsterdam: John Benjamin}, pp. 265--276, 2011.

\bibitem{zhao2017bivariate}
H.~Zhao and J.~Chen, ``Bivariate separable-dimension glyphs can improve visual
  analysis of holistic features,'' \emph{arXiv:
  https://arxiv.org/abs/1712.02333v1}, 2017.

\bibitem{kindlmann2002face}
G.~Kindlmann, E.~Reinhard, and S.~Creem, ``Face-based luminance matching for
  perceptual colormap generation,'' in \emph{Proceedings of the Conference on
  Visualization}, 2002, pp. 299--306.

\bibitem{hu2014interactive}
G.~Hu, Z.~Pan, M.~Zhang, D.~Chen, W.~Yang, and J.~Chen, ``An interactive method
  for generating harmonious color schemes,'' \emph{Color Research \&
  Application}, vol.~39, no.~1, pp. 70--78, 2014.

\bibitem{gramazio2017colorgorical}
C.~C. Gramazio, D.~H. Laidlaw, and K.~B. Schloss, ``Colorgorical: Creating
  discriminable and preferable color palettes for information visualization,''
  \emph{IEEE Transactions on Visualization and Computer Graphics}, vol.~23,
  no.~1, pp. 521--530, 2017.

\bibitem{bujack2017good}
R.~Bujack, T.~L. Turton, F.~Samsel, C.~Ware, D.~H. Rogers, and J.~Ahrens, ``The
  good, the bad, and the ugly: A theoretical framework for the assessment of
  continuous colormaps,'' \emph{IEEE Transactions on Visualization and Computer
  Graphics}, vol.~1, no.~1, pp. 923 -- 933, 2018.

\bibitem{szafir2017modeling}
D.~A. Szafir, ``Modeling color difference for visualization design,''
  \emph{IEEE Transactions on Visualization and Computer Graphics}, vol.~24,
  no.~1, pp. 392--401, 2018.

\bibitem{trumbo1981theory}
B.~E. Trumbo, ``A theory for coloring bivariate statistical maps,'' \emph{The
  American Statistician}, vol.~35, no.~4, pp. 220--226, 1981.

\bibitem{ware2017evaluating}
C.~Ware, T.~L. Turton, F.~Samsel, R.~Bujack, D.~H. Rogers, K.~Lawonn, N.~Smit,
  and D.~Cunningham, ``Evaluating the perceptual uniformity of color sequences
  for feature discrimination,'' in \emph{EuroVis Workshop on Reproducibility,
  Verification, and Validation in Visualization (EuroRV3). The Eurographics
  Association}, 2017.

\bibitem{penney2012effects}
D.~Penney, J.~Chen, and D.~H. Laidlaw, ``Effects of illumination, texture, and
  motion on task performance in 3d tensor-field streamtube visualizations,'' in
  \emph{IEEE Pacific Visualization Symposium}, 2012, pp. 97--104.

\bibitem{zhao2017validation}
H.~Zhao, G.~W. Bryant, W.~Griffin, J.~E. Terrill, and J.~Chen, ``Validation of
  splitvectors encoding for quantitative visualization of large-magnitude-range
  vector fields,'' \emph{IEEE Transactions on Visualization and Computer
  Graphics}, vol.~23, no.~6, pp. 1691--1705, 2017.

\bibitem{ritter2006real}
F.~Ritter, C.~Hansen, V.~Dicken, O.~Konrad, B.~Preim, and H.-O. Peitgen,
  ``Real-time illustration of vascular structures,'' \emph{IEEE Transactions on
  Visualization and Computer Graphics}, vol.~12, no.~5, pp. 877--884, 2006.

\bibitem{svetachov2010dti}
P.~Svetachov, M.~H. Everts, and T.~Isenberg, ``{DTI} in context: illustrating
  brain fiber tracts in situ,'' in \emph{Computer Graphics Forum}, vol.~29,
  no.~3, 2010, pp. 1023--1032.

\bibitem{acevedo2006subjective}
D.~Acevedo and D.~Laidlaw, ``Subjective quantification of perceptual
  interactions among some 2d scientific visualization methods,'' \emph{IEEE
  Transactions on Visualization and Computer Graphics}, vol.~12, no.~5, pp.
  1133--1140, 2006.

\bibitem{borkin2011evaluation}
M.~Borkin, K.~Gajos, A.~Peters, D.~Mitsouras, S.~Melchionna, F.~Rybicki,
  C.~Feldman, and H.~Pfister, ``Evaluation of artery visualizations for heart
  disease diagnosis,'' \emph{IEEE Transactions on Visualization and Computer
  Graphics}, vol.~17, no.~12, pp. 2479--2488, 2011.

\bibitem{chen2012effects}
J.~Chen, H.~Cai, A.~P. Auchus, and D.~H. Laidlaw, ``Effects of stereo and
  screen size on the legibility of three-dimensional streamtube
  visualization,'' \emph{IEEE Transactions on Visualization and Computer
  Graphics}, vol.~18, no.~12, pp. 2130--2139, 2012.

\bibitem{kindlmann1999hue}
G.~Kindlmann and D.~Weinstein, ``Hue-balls and lit-tensors for direct volume
  rendering of diffusion tensor fields,'' in \emph{Proceedings of the
  Conference on Visualization}, 1999, pp. 183--189.

\bibitem{maule2014getting}
J.~Maule, C.~Witzel, and A.~Franklin, ``Getting the gist of multiple hues:
  metric and categorical effects on ensemble perception of hue,'' \emph{Journal
  of the Optical Society of America. A, Optics, Image Science, and Vision},
  vol.~31, no.~4, pp. A93--102, 2014.

\bibitem{munzner2009nested}
T.~Munzner, ``A nested model for visualization design and validation,''
  \emph{IEEE Transactions on Visualization and Computer Graphics}, vol.~15,
  no.~6, pp. 921--928, 2009.

\bibitem{mori2006principles}
S.~Mori and J.~Zhang, ``Principles of diffusion tensor imaging and its
  applications to basic neuroscience research,'' \emph{Neuron}, vol.~51, no.~5,
  pp. 527--539, 2006.

\bibitem{zhang2017overview}
C.~Zhang, M.~Caan, T.~H{\"o}llt, E.~Eisemann, and A.~Vilanova, ``Overview+
  detail visualization for ensembles of diffusion tensors,'' in \emph{Computer
  Graphics Forum}, vol.~36, no.~3, 2017, pp. 121--132.

\bibitem{isenberg2013systematic}
T.~Isenberg, P.~Isenberg, J.~Chen, M.~Sedlmair, and T.~M{\"o}ller, ``A
  systematic review on the practice of evaluating visualization,'' \emph{IEEE
  Transactions on Visualization and Computer Graphics}, vol.~19, no.~12, pp.
  2818--2827, 2013.

\bibitem{preim2016survey}
B.~Preim, A.~Baer, D.~Cunningham, T.~Isenberg, and T.~Ropinski, ``A survey of
  perceptually motivated {3D} visualization of medical image data,'' in
  \emph{Computer Graphics Forum}, vol.~35, no.~3.\hskip 1em plus 0.5em minus
  0.4em\relax Wiley Online Library, 2016, pp. 501--525.

\bibitem{kochunov2016heterochronicity}
P.~Kochunov, H.~Ganjgahi, A.~Winkler, S.~Kelly, D.~K. Shukla, X.~Du,
  N.~Jahanshad, L.~Rowland, H.~Sampath, B.~Patel, P.~O'Donnell, Z.~Xie, S.~A.
  Paciga, C.~R. Schubert, J.~Chen, G.~Zhang, P.~M. Thompson, T.~E. Nichols, and
  H.~L. Elliot, ``Heterochronicity of white matter development and aging
  explains regional patient control differences in schizophrenia,'' \emph{Human
  Brain Mapping}, vol.~37, no.~12, pp. 4673--4688, 2016.

\bibitem{zhang2003visualizing}
S.~Zhang, C.~Demiralp, and D.~H. Laidlaw, ``Visualizing diffusion tensor {MR}
  images using streamtubes and streamsurfaces,'' \emph{IEEE Transactions on
  Visualization and Computer Graphics}, vol.~9, no.~4, pp. 454--462, 2003.

\bibitem{jiang2006dtistudio}
H.~Jiang, P.~C. van Zijl, J.~Kim, G.~D. Pearlson, and S.~Mori, ``{DtiStudio}:
  resource program for diffusion tensor computation and fiber bundle
  tracking,'' \emph{Computer Methods and Programs in Biomedicine}, vol.~81,
  no.~2, pp. 106--116, 2006.

\bibitem{pieper20043d}
S.~Pieper, M.~Halle, and R.~Kikinis, ``{3D} slicer,'' in \emph{IEEE
  International Symposium on Biomedical Imaging: Nano to Macro}, 2004, pp.
  632--635.

\bibitem{borgo2014order}
R.~Borgo, J.~Dearden, and M.~W. Jones, ``Order of magnitude markers: An
  empirical study on large magnitude number detection,'' \emph{IEEE
  Transactions on Visualization and Computer Graphics}, vol.~20, no.~12, pp.
  2261--2270, 2014.

\bibitem{sharma2005ciede2000}
G.~Sharma, W.~Wu, and E.~N. Dalal, ``The {CIEDE2000} color-difference formula:
  Implementation notes, supplementary test data, and mathematical
  observations,'' \emph{Color Research \& Application}, vol.~30, no.~1, pp.
  21--30, 2005.

\bibitem{brun2003coloring}
A.~Brun, H.-J. Park, H.~Knutsson, and C.-F. Westin, ``Coloring of {DT-MRI}
  fiber traces using laplacian eigenmaps,'' in \emph{International Conference
  on Computer Aided Systems Theory}, 2003, pp. 518--529.

\bibitem{belkin2003laplacian}
M.~Belkin and P.~Niyogi, ``Laplacian eigenmaps for dimensionality reduction and
  data representation,'' \emph{Neural Computation}, vol.~15, no.~6, pp.
  1373--1396, 2003.

\bibitem{cohen1988statistical}
J.~Cohen, \emph{Statistical power analysis for the behavioral sciences}.\hskip
  1em plus 0.5em minus 0.4em\relax New York: Academic Press, 1988.

\bibitem{cox2013visualizing}
J.~Cox and M.~Lindell, ``Visualizing uncertainty in predicted hurricane
  tracks,'' \emph{International Journal for Uncertainty Quantification},
  vol.~3, no.~2, 2013.

\bibitem{chen2009visual}
J.~Chen, M.~Kostandov, I.~Pivkin, D.~Riskin, D.~Willis, S.~Swartz, and
  D.~Laidlaw, ``Visual analysis of dimensionality reduction for exploring bat
  flight kinematics in a virtual environment,'' in \emph{Proceedings of the
  15th Joint virtual reality Eurographics conference on Virtual
  Environments}.\hskip 1em plus 0.5em minus 0.4em\relax Eurographics
  Association, 2009, pp. 77--84.

\bibitem{andrienko2016leveraging}
N.~Andrienko, G.~Andrienko, and S.~Rinzivillo, ``Leveraging spatial abstraction
  in traffic analysis and forecasting with visual analytics,''
  \emph{Information Systems}, vol.~57, pp. 172--194, 2016.

\bibitem{liu20122d}
Z.~Liu, S.~Cai, J.~E. Swan, R.~J. Moorhead, J.~P. Martin, and T.~Jankun-Kelly,
  ``A {2D} flow visualization user study using explicit flow synthesis and
  implicit task design,'' \emph{IEEE Transactions on Visualization and Computer
  Graphics}, vol.~18, no.~5, pp. 783--796, 2012.

\bibitem{peng2012mesh}
Z.~Peng, E.~Grundy, R.~S. Laramee, G.~Chen, and N.~Croft, ``Mesh-driven vector
  field clustering and visualization: An image-based approach,'' \emph{IEEE
  Transactions on Visualization and Computer Graphics}, vol.~18, no.~2, pp.
  283--298, 2012.

\bibitem{li2011visbubbles}
G.~Li, A.~C. Bragdon, Z.~Pan, M.~Zhang, S.~M. Swartz, D.~H. Laidlaw, C.~Zhang,
  H.~Liu, and J.~Chen, ``Visbubbles: a workflow-driven framework for scientific
  data analysis of time-varying biological datasets,'' in \emph{ACM SIGGRAPH
  Asia Posters}, 2011, p.~27.

\bibitem{ragan2016characterizing}
E.~D. Ragan, A.~Endert, J.~Sanyal, and J.~Chen, ``Characterizing provenance in
  visualization and data analysis: an organizational framework of provenance
  types and purposes,'' \emph{IEEE Transactions on Visualization and Computer
  Graphics}, vol.~22, no.~1, pp. 31--40, 2016.

\bibitem{rogowitz2001blair}
B.~E. Rogowitz and A.~D. Kalvin, ``The ``which blair project'': a quick visual
  method for evaluating perceptual color maps,'' in \emph{Proceedings of the
  Conference on Visualization}, 2001, pp. 183--556.

\end{thebibliography}
%



%






\begin{IEEEbiography}
	[{\includegraphics[width=1in,height=1.25in,clip,keepaspectratio]{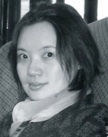}}]
	{Jian Chen} received the PhD degree in Computer Science from Virginia Polytechnic Institute and State University (Virginia Tech). She did her post-doctoral work in the Department of Computer Science at Brown University. She is an Associate Professor in Computer Science and Electrical Engineering at The Ohio State University where she directs the Interactive Visual Computing Laboratory (IVCL). Her research interests include design and evaluation of visualization techniques and virtual reality. She is a member of the IEEE and the IEEE Computer Society.
\end{IEEEbiography}

\begin{IEEEbiography}
	[{\includegraphics[width=1in,height=1.25in,clip,keepaspectratio]{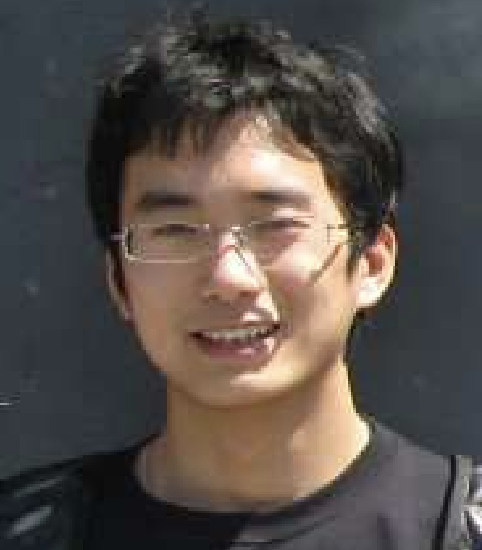}}]
	{Guohao Zhang} is a PhD student in the Department of Computer Science and Electrical Engineering at University of Maryland, Baltimore County. He received his B.E. degree in Engineering Physics from Tsinghua University in 2012. His research interests include design and evaluation of visualization techniques and 3D visualizations. He is a student member of IEEE.
\end{IEEEbiography}

\begin{IEEEbiography}
	[{\includegraphics[width=1in,height=1.25in,clip,keepaspectratio]{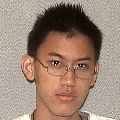}}]
	{Wesley Chiou}
	 is an undergraduate student in the Department of Computer Science and Electrical Engineering at University of Maryland, Baltimore County. His research interest is human-computer interaction and visualization.
\end{IEEEbiography}

\begin{IEEEbiography}
	[{\includegraphics[width=1in,height=1.25in,clip,keepaspectratio]{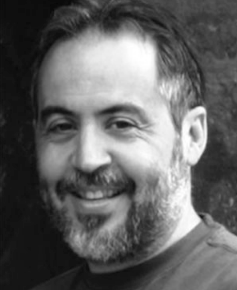}}]
	{David H. Laidlaw}
	 received the PhD degree in
computer science from the California Institute of
Technology, where he also did post-doctoral work
in the Division of Biology. He is a professor in the
Computer Science Department at Brown University.
His research centers on applications of visualization,
modeling, computer graphics, and
computer science to other scientific disciplines.
He is a fellow of IEEE.
\end{IEEEbiography}

\begin{IEEEbiography}
	[{\includegraphics[width=1in,height=1.25in,clip,keepaspectratio]{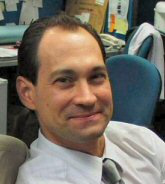}}]
	{Alexander P. Auchus}
	Dr. Alexander P. Auchus holds degrees from Johns Hopkins University and from Washington University in St. Louis.  He is an elected fellow of the American Neurological Association, the American Academy of Neurology, and the American Geriatrics Society. He has served on the faculty of Emory University, Case Western Reserve University, and University of Tennessee.  His present position is Professor and McCarty Chair of Neurology at the University of Mississippi Medical Center. Dr. Auchus's research interests are in neuroimaging biomarkers for Alzheimer's disease and other dementias.
\end{IEEEbiography}







\end{document}